\documentclass{sig-alternate-05-2015}

\usepackage[scaled=0.85]{beramono}
\usepackage[T1]{fontenc}
\usepackage{microtype}
\usepackage{graphicx}
\usepackage{balance}  
\usepackage{xspace}
\usepackage{textcomp}
\usepackage{subcaption}
\usepackage{enumitem}
\usepackage{color}
\usepackage{siunitx}
\usepackage{booktabs}
\usepackage{enumitem}
\usepackage{pifont}

\usepackage{tikz}
\usetikzlibrary{%
 arrows.meta,%
 calc,%
 fit,%
 patterns,%
 plotmarks,%
 shapes.geometric,%
 shapes.misc,%
 shapes.symbols,%
 shapes.arrows,%
 shapes.callouts,%
 shapes.multipart,%
 backgrounds,%
 chains,%
 topaths,%
 trees,%
 matrix,%
 folding,%
 fadings,%
 through,%
 positioning,%
 quotes,%
 scopes,%
 decorations.fractals,%
 decorations.shapes,%
 decorations.text,%
 decorations.pathmorphing,%
 decorations.pathreplacing,%
 decorations.footprints,%
 decorations.markings,%
 shadows
}

\begin{document}

\CopyrightYear{2017} 
\setcopyright{acmlicensed}
\doi{http://dx.doi.org/10.1145/3035918.3064046}

%

\title{Living in Parallel Realities -- Co-Existing Schema Versions with a Bidirectional Database Evolution Language
}

\numberofauthors{5} %

\author{
\alignauthor
Kai Herrmann\titlenote{Funded by the German Research Foundation (DFG) within the RoSI Research Training Group (GRK 1907).\newline Project: \url{wwwdb.inf.tu-dresden.de/research-projects/inverda}}\\
       \affaddr{Technische Universit\"{a}t Dresden, Germany}\\
       \email{kai.herrmann@}
       \email{tu-dresden.de}
\and
\alignauthor
Hannes Voigt\\
       \affaddr{Technische Universit\"{a}t Dresden, Germany}\\
       \email{hannes.voigt@}
       \email{tu-dresden.de}
\and
\alignauthor
Andreas Behrend\\
       \affaddr{Universit\"{a}t Bonn,}\\
       \affaddr{Germany}\\
       \email{behrend@}
       \email{cs.uni-bonn.de}
\and
\alignauthor
Jonas Rausch\\
       \affaddr{Technische Universit\"{a}t Dresden, Germany}\\
       \email{jonas.rausch@}
       \email{tu-dresden.de}
\and
\alignauthor
Wolfgang Lehner\\
       \affaddr{Technische Universit\"{a}t Dresden, Germany}\\
       \email{wolfgang.lehner@}
       \email{tu-dresden.de}
}

\date{\today}

\maketitle

\newcommand{\PARTITION}{SPLIT\xspace}
\newcommand{\Partition}{Split\xspace}
\newcommand{\partition}{split\xspace}

\newcommand{\tickYes}{{\color[rgb]{0,.4,0}\checkmark}}
\newcommand{\tickNo}{{\color[rgb]{.6,0,0}\ding{55}}}
\newcommand{\tickIgnore}{\tickNo}
\newcommand{\tickYesGrey}{{\color[rgb]{0.4,.4,0.4}\checkmark}}
\newcommand{\tickNoGrey}{{\color[rgb]{.4,0.4,0.4}\ding{55}}}
\newcommand{\tickIgnoreGrey}{\tickNoGrey}

\newcommand{\vldb}[2]{#1}

\newcommand{\codel}{\textsc{CoDEL}\xspace}
\newcommand{\indel}{\textsc{BiDEL}\xspace}
\newcommand{\inverda}{\textsc{InVerDa}\xspace}
\newcommand{\inverdaHeading}{I{\normalfont \textbf{N}}V{\normalfont \textbf{ER}}D{\normalfont \textbf{A}}\xspace}
\newcommand{\indelHeading}{B{\normalfont \textbf{I}}DEL$\;$\xspace}
\newcommand{\indelHeadings}{B{\normalfont \textbf{I}}DEL's$\;$\xspace}

\newcommand{\sql}[1]{\texttt{\footnotesize{#1}}}
\newcommand{\sqll}[1]{\par\hangindent=\parindent\vspace{.5ex}\sql{#1}\vspace{.5ex}\par\noindent}
\newcommand{\sqlc}[1]{\textquotesingle#1\textquotesingle}
\newcommand{\sqlk}[1]{\textbf{#1}}

\newcommand{\pk}{\ensuremath{p}\xspace}
\newcommand{\getX}[1]{\ensuremath{\gamma_{#1}}\xspace}
\newcommand{\getSource}{\getX{src}}
\newcommand{\getTarget}{\getX{tgt}}
\newcommand{\Tdata}{\ensuremath{T_{D}}\xspace}
\newcommand{\Rdata}{\ensuremath{R_{D}}\xspace}
\newcommand{\RPdata}{\ensuremath{R'_{D}}\xspace}
\newcommand{\Sdata}{\ensuremath{S_{D}}\xspace}
\newcommand{\Tnew}{\ensuremath{T_{new}}\xspace}

\newcommand{\cR}{\ensuremath{c_R}\xspace}
\newcommand{\cS}{\ensuremath{c_S}\xspace}
\newcommand{\cT}{\ensuremath{c_T}\xspace}
\newcommand{\ds}{\ensuremath{D_{src}}\xspace}
\newcommand{\dt}{\ensuremath{D_{tgt}}\xspace}
\newcommand{\cond}{\ensuremath{c}\xspace}

\newcommand{\taskA}{\sql{Task}}
\newcommand{\authorA}{\sql{Author}}
\newcommand{\todos}{\sql{Todo}}
\newcommand{\taskR}[1]{\sql{Task-#1}}
\newcommand{\authorR}[1]{\sql{Author-#1}}
\newcommand{\todoR}[1]{\sql{Todo-#1}}
\newcommand{\cTask}{\sql{task}}
\newcommand{\cAuthor}{\sql{author}}
\newcommand{\cPrio}{\sql{prio}}
\newcommand{\cName}{\sql{name}}
\newcommand{\tasky}{TasKy}
\newcommand{\taskyy}{TasKy2}
\newcommand{\dodo}{Do!}
\newcommand{\elt}[1]{{\sffamily#1}}
\newcommand{\taskyX}{\elt{\tasky}\xspace}
\newcommand{\taskyyX}{\elt{\taskyy}\xspace}
\newcommand{\dodoX}{\elt{\dodo}\xspace}
\newcommand{\tbl}[2]{\sql{#1(#2)}}

\newcommand{\tpchA}{$TPCH_0$\xspace}
\newcommand{\tpchB}{$TPCH_1$\xspace}
\newcommand{\tpchC}{$TPCH_2$\xspace}

\newcommand{\set}[1]{\ensuremath{\left\{#1\right\}}}
\newcommand{\lst}[1]{\ensuremath{\left[#1\right]}}
\newcommand{\brk}[1]{\ensuremath{\left(#1\right)}}
\newcommand{\abs}[1]{\ensuremath{\left|#1\right|}}
\newcommand{\pst}[1]{\ensuremath{\mathcal{P}{\brk{#1}}}}

\newcommand{\op}[1]{\ensuremath{\texttt{#1}}}
\newcommand{\rvars}[1]{\ensuremath{\op{vars}(#1)}}
\newcommand{\rpred}[1]{\ensuremath{\op{pred}(#1)}}
\newcommand{\rhead}[1]{\ensuremath{\op{head}(#1)}}
\newcommand{\rbody}[1]{\ensuremath{\op{body}(#1)}}
\newcommand{\renm}[1]{\ensuremath{\op{rn}(#1)}}
\newcommand{\ruleset}[1]{\ensuremath{\mathcal{#1}}}
\newcommand{\srctabs}[1]{\ensuremath{\op{src}(#1)}}
\newcommand{\trgtabs}[1]{\ensuremath{\op{tgt}(#1)}}
\newcommand{\insmo}[1]{\ensuremath{\op{in}(#1)}}
\newcommand{\outsmo}[1]{\ensuremath{\op{out}(#1)}}

\newtheorem{theorem}{Theorem}
\newtheorem{lemma}[theorem]{Lemma}



\newcommand{\tab}[1][black]{
    \draw[thick,#1,fill=white] (0,0) rectangle (3mm,3mm);
    \draw[thick,#1] (0,2mm) to (3mm,2mm);
    \draw[#1] (0,1mm) to (3mm,1mm);
    \draw[#1] (1mm,0) to (1mm,3mm);
    \draw[#1] (2mm,0) to (2mm,3mm);
}
\newcommand{\view}{
    \node(t)[inner sep=0,outer sep=0] {\tikz\tab[darkgray];};
    \coordinate (g mid)          at ($(t.south west)+(.8mm,0.7mm)$);
    \coordinate (g left glas)    at ($(g mid)+(-2mm,0)$);
    \coordinate (g right glas)   at ($(g mid)+(2mm,0)$);
    \coordinate (g left side)    at ($(g mid)+(-0.5mm,2mm)$);
    \coordinate (g right side)   at ($(g mid)+(3mm,2mm)$);
    \draw[thick,black,fill=lightgray,fill opacity=0.5] ($(g left glas)+(0.75mm,0)$) circle [radius=0.75mm];
    \draw[thick,black,fill=lightgray,fill opacity=0.5] ($(g right glas)+(-0.75mm,0)$) circle [radius=0.75mm];
    \draw[thick,black] ($(g left glas)+(1.5mm,0)$)--($(g right glas)+(-1.5mm,0)$);
    \draw[thick,black,-{Hooks[right]}] (g left glas) to (g left side);
    \draw[thick,black,-{Hooks[right]}] (g right glas) to (g right side);
}
\newcommand{\gear}{
    \draw[fill=lightgray] (1.5mm,0) foreach \x [evaluate=\x as \y using \x+30] in {10,55,100,145,190,235,280,325} {
        arc[radius=1.5mm,start angle=\x,delta angle=30] -- ([turn]90:0.3mm)
        arc[radius=1.2mm,start angle=\y,delta angle=15] -- ([turn]270:0.3mm) 
    };
    \draw[fill=white] (0mm,-0.25mm) circle [radius=0.5mm];
}
\newcommand{\viewtrigger}{
    \node(v)[inner sep=0,outer sep=0] {\tikz\view;};
    \node(t)[inner sep=0,outer sep=0,anchor=center] at ($(v.south east)+(0mm,0.2mm)$) {\tikz\gear;};
}

\begin{abstract}
We introduce end-to-end support of co-existing schema versions within one database.
While it is state of the art to run multiple versions of a continuously developed application concurrently, it is hard to do the same for databases.
In order to keep multiple co-existing schema versions alive---which are all accessing the same data set---developers usually employ handwritten delta code (e.g. views and triggers in SQL).
This delta code is hard to write and hard to maintain: if a database administrator decides to adapt the physical table schema, all handwritten delta code needs to be adapted as well, which is expensive and error-prone in practice.
In this paper, we present \inverda: developers use the simple bidirectional database evolution language \indel, which carries enough information to generate all delta code automatically.
Without additional effort, new schema versions become immediately accessible and data changes in any version are visible in all schema versions at the same time.
\inverda also allows for easily changing the physical table design without affecting the availability of co-existing schema versions.
This greatly increases robustness (orders of magnitude less lines of code) and allows for significant performance optimization.
A main contribution is the formal evaluation that each schema version acts like a common full-fledged database schema independently of the chosen physical table design.

\end{abstract}

%
%


%
%

%
%
\begin{sloppypar}
\printccsdesc
\end{sloppypar}



\section{Introduction}

Database management systems (DBMSes) lack proper support for co-existing schema versions within the same database.
With today's realities in information system development and deployment---namely agile development methods, code refactoring, short release cycles, stepwise deployment, varying update adoption time, legacy system support, etc.---such support becomes increasingly desirable.
Tools such as GIT, SVN, and Maven allow to maintain multiple versions of one application and deploy several of these versions concurrently.
The same is hard for database systems, though.
In enterprise information systems, databases feed hundreds of subsystems across the company domain, connecting decades-old legacy systems with brand new web front ends or innovative analytics pipelines~\cite{Brodie2010}.
These subsystems are typically run by different stakeholders, with different development cycles and different upgrade constraints.
Typically, adopting changes to a database schema---if even possible on the client-side---will stretch out over time.
Hence, the database schema versions these subsystems expect have to be kept alive on the database side to continuously serve them.
Co-existing schema versions are an actual challenge in information systems developers and database administrators (DBAs) have to cope with. 

Unfortunately, current DBMSes do not support co-existing schema versions properly and essentially force developers to migrate a database completely in one haul to a new schema version.
Keeping old schema versions alive to continue serving all database clients independent of their adoption time is costly.
Before and after a one-haul migration, manually written and maintained \emph{delta code} is required. 
Delta code is any implementation of propagation logic to run an application with a schema version differing from the version used by the database.
Delta code may comprise view and trigger definitions in the database, propagation code in the database access layer of the application, or ETL jobs for update propagation for a database replicated in different schema versions.
Data migration accounted for \SI{31}{\percent} of IT project budgets in 2011 and co-existing schema versions take an important part in that~\cite{Howard2011}.
In short, handling co-existing schema versions is very costly and error-prone and forces developers into less agility, longer release cycles, riskier big-bang migration, etc.

\newcommand*\rot{\rotatebox{90}}
\newcommand{\spaceTab}{3mm}
\begin{table}
\centering\scriptsize
\begin{tabular}{lc@{\hspace{\spaceTab}}c@{\hspace{\spaceTab}}c@{\hspace{\spaceTab}}c@{\hspace{\spaceTab}}c@{\hspace{\spaceTab}}c@{\hspace{\spaceTab}}c@{\hspace{1mm}}c@{\hspace{0mm}}}
\toprule
\mbox{
\begin{tabular}{l l}
\tickYes \tickYesGrey& supported\\
\tickNo\ \tickNoGrey& not supported\\[2.27cm]

\end{tabular}
}\vspace{-1.4cm}
&\rot{SQL}
&\rot{Model Mngt}
&\rot{PRISM}
&\rot{PRIMA}
&\rot{CoDEL}
&\rot{Sym.\ Lenses}
&\rot{\textbf{\indel}}
&\rot{\textbf{\inverda}}\\
\midrule
\hspace{-1.5 mm}Database Evolution Language&\tickNo&\tickNo&\tickYes&\tickYes&\tickYes&\tickNo&\multicolumn{2}{c}{\tickYes}\\
\hspace{-1.5 mm}Relationally Complete&\tickYes&\tickYes&\tickNo&\tickNo&\tickYes&\tickIgnore&\multicolumn{2}{c}{\tickYes}\\
\hspace{-1.5 mm}Co-Existing Schema Versions&\tickIgnore&\tickNo&\tickNo&(\tickYes)&\tickNo&\tickYes&\multicolumn{2}{c}{\tickYes}\\
\hspace{-1mm}- \emph{Forward Query Rewriting}&\tickIgnoreGrey&\tickYesGrey&\tickYesGrey&\tickYesGrey&\tickNoGrey&\tickYesGrey&\multicolumn{2}{c}{\tickYesGrey}\\
\hspace{-1mm}- \emph{Backward Query Rewriting}&\tickIgnoreGrey&\tickNoGrey&\tickNoGrey&\tickNoGrey&\tickNoGrey&\tickYesGrey&\multicolumn{2}{c}{\tickYesGrey}\\
\hspace{-1mm}- \emph{Forward Migration}&\tickIgnoreGrey&\tickYesGrey&\tickYesGrey&\tickYesGrey&\tickYesGrey&\tickYesGrey&\multicolumn{2}{c}{\tickYesGrey}\\
\hspace{-1mm}- \emph{Backward Migration}&\tickIgnoreGrey&\tickNoGrey&\tickNoGrey&\tickNoGrey&\tickNoGrey&\tickYesGrey&\multicolumn{2}{c}{\tickYesGrey}\\
\hspace{-1.5 mm}Guaranteed Bidirectionality&\tickIgnore&\tickNo&\tickNo&\tickNo&\tickNo&\tickYes&\multicolumn{2}{c}{\tickYes}\\
\bottomrule
\end{tabular}\\[-2mm]
\caption{Contribution and Distinction from Related Work.}
\vspace{-2.5mm}
\label{tab:rw}
\end{table}

In this paper, we present the Database Evolution Language (DEL) \indel (\underline{Bi}directional \underline{DEL}).
\indel provides simple but powerful bidirectional Schema Modification Operations (SMOs) that define the evolution of both the schema and the data from one schema version to a new one. 
\textbf{Bidirectionality} is the unique feature of \indel's SMOs and the central concept to facilitate co-existing schema versions by allowing full access propagation between all schema versions.
\indel builds upon SMOs of existing \emph{monodirectional} DELs~\cite{Curino2012,HVBL15} and extends them, thus they become bidirectional.\
With a \textbf{formal evaluation of bidirectionality}, we guarantee that the propagation of read and write operations on any schema version to all other schema versions works always correctly. 

As a proof of concept for \indel, we present \inverda~\cite{Herrmann2016a} (\underline{In}tegrated \underline{Ver}sioning of \underline{Da}tabases).
\inverda is an extension to relational DBMSes to enable true support for co-existing schema versions based on \indel.
It allows a single database to have multiple co-existing schema versions that are all accessible at the same time. 
Specifically, \inverda introduces two powerful functionalities to a DBMS.

For application developers, \inverda offers a \textbf{Database Evolution Operation} that executes a \indel-specified evolution from an existing schema version to a new version.
New schema versions become immediately available.
Applications can read and write data through any schema version; writes in one version are reflected in all other versions.
Each schema version itself appears to the user like a full-fledged single-schema database.
The \inverda Database Evolution Operation generates all necessary delta code, more precisely views and triggers within the database, so applications can read and write those views as usual, i.e., the complete access propagation between all schema versions is implemented with one click of a button.
\inverda greatly simplifies evolution tasks, since the developer has to write a \indel script only.


For DBAs, \inverda offers a single-line \textbf{Database Migration Operation} to configure the primarily materialized schema version.
If the workload mix changes, because e.g. most client applications use a new version, the DBA easily changes the materialization without affecting the availability of any schema version and without a developer being involved.
Thanks to \indel's bidirectionality, \inverda already has all required information to migrate the affected data and to regenerate all delta code---not a single line of code is required from the developer.
Without \inverda such an optimization requires to rewrite all affected delta code manually.

\inverda generates views and triggers as delta code.
Other implementations of \indel are very well imaginable, e.g. generation of ETL jobs or application-side propagation logic.
However, we are convinced that a functional extension to database systems is the most appealing approach. 

\indel and \inverda are not the first attempts to support schema evolution.
For practitioners, valuable tools, such as Liquibase, Rails Migrations, and DBmaestro Teamwork, help to manage schema versions outside the DBMS and generate SQL scripts for migrating to a new schema version.
They mitigate data migration costs, but focus on schema evolution and support for co-existing schemas is very limited.

For research, Table~\ref{tab:rw} classifies related work regarding support for co-existing schema versions and highlights the contributions of \indel and \inverda.
Meta Model Management helps handling multiple schema versions ``after the fact'' by allowing to match, diff, and merge existing schemas to derive mappings between these schemas~\cite{Bernstein2007b}. 
The derived mappings are expressed with relational algebra and can be used to rewrite old queries or to migrate data forwards---not backwards, though.
In contrast, the inspiring PRISM/PRISM++ proposes to let the developer specify the evolution with SMOs ``before the fact'' to a derived new schema version~\cite{Curino2012}.
PRISM proposes an intuitive and declarative DEL that documents the evolution and implicitly allows migrating data forwards and rewriting queries from old to new schema versions.
As an extension, PRIMA~\cite{Moon2009} takes a first step towards co-existing schema versions by propagating write operations forwards and read operations backwards along the schema version history, but not vice versa.
CoDEL~\cite{HVBL15} slightly extends the PRISM DEL to a relationally complete DEL.
\indel now extends CoDEL to be bidirectional while maintaining relational completeness.
According to our evaluation, \indel SMOs are just as compact as PRISM SMOs and thereby orders of magnitude shorter than SQL.
To our best knowledge, there is no bidirectional DEL nor a comprehensive solution for co-existing schema versions so far.
Symmetric relational lenses~\cite{Hofmann2011} define abstract formal conditions that mapping functions need to satisfy in order to be bidirectional.
However, they do not specify any concrete mapping as specific as the semantics of SMOs required to form a DEL.
To our best knowledge \indel is the first DEL with SMOs intentionally designed and formally evaluated to fulfill the symmetric lense conditions and \inverda is the first approach to implement such a bidirectional DEL in a relational DBMS.
In sum, the contributions of this paper are:

\begin{description}[leftmargin=6mm] 
\item[Formally evaluated, bidirectional DEL:]
We introduce syntax and semantics of \indel and formally validate its bidirectionality by showing that its SMOs fulfill the symmetric lense conditions.
\indel greatly supports developers and DBAs.
In our examples, \indel requires significantly (up to \SI{359}{x}) less code than evolutions and migrations manually written in SQL.
(Sections~\ref{sec:singlesmo} and~\ref{sec:bidirectionality})
\item[Co-existing schema versions:] 
\inverda's Database Evolution Operation automatically generates delta code from \indel-specified schema evolutions to allow reads and writes on all co-existing schema versions, each providing an individual view on the same dataset.
The delta code generation is really fast~($\SI{<1}{\second}$) and query performance is comparable to hand-written delta code.
(Section~\ref{sec:generation})
\item[Logical data independence:] 
\inverda's Database Migration Operation makes manual schema migrations obsolete.
It triggers the physical data movement as well as the adaptation of all involved delta code and allows the DBA to optimize the physical table schema of the database independently from the schema evolution, which yields significant performance improvements.
(Section~\ref{sec:migration})
\end{description}\newpage



We introduce \inverda from a user perspective in Section~\ref{sec:user} and discuss its architecture in Section~\ref{sec:architecture}.
In Section~\ref{sec:singlesmo}, we present \indel and formally evaluate its bidirectionality in Section~\ref{sec:bidirectionality}.
In Sections~\ref{sec:generation} and~\ref{sec:migration}, we sketch how to generate delta code and change the materialization schema.
Finally, we evaluate \inverda in Section~\ref{sec:evaluation}, discuss related work in Section~\ref{sec:rw}, and conclude the paper in Section~\ref{sec:conclusion}.

\section{User Perspective on \inverdaHeading}\label{sec:user}
\begin{figure}
\centering
\includegraphics[width=\columnwidth, page=8, trim=0mm 99mm 105mm 0mm, clip=true]{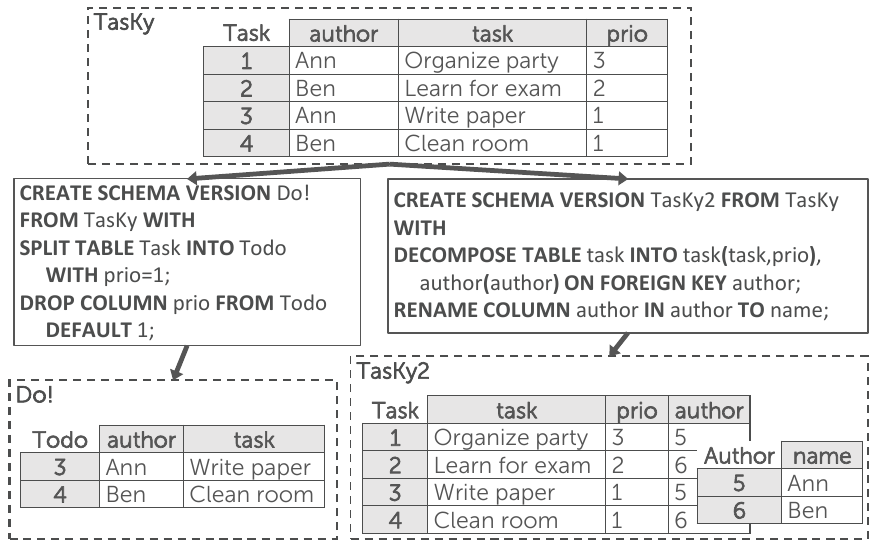}\\[-2mm]
\caption{\taskyX Example.}
\vspace{-1mm}
\label{fig:example}
\end{figure}

In the following, we introduce the co-existing schema version support by \inverda using the example of a simple task management system called \taskyX (cf. Figure~\ref{fig:example}).
\taskyX is a desktop application which is backed by a central database.
It allows users creating new tasks as well as listing, updating, and deleting them. %
Each task has an author and a priority.
Tasks with priority 1 are the most urgent ones. %
The first release of \taskyX stores all its data in a single table \tbl{\taskA}{\cAuthor,\cTask,\cPrio}.
\taskyX has productive go-live and users begin to feed the database with their tasks.

\textbf{Developer: }
\taskyX gets widely accepted and after some weeks it is extended by a third party phone app called \dodoX to list most urgent tasks.
However \dodoX expects a different database schema than \taskyX is using.
The \dodoX schema consists of a table \tbl{\todos}{\cAuthor,\cTask} containing only tasks of priority 1.
Obviously, the initial schema version needs to stay alive for \taskyX, which is broadly installed.
\inverda greatly simplifies the job as it handles all the necessary delta code for the developer.
The developer merely executes the \indel script for \dodoX as shown in Figure~\ref{fig:example}, which instructs \inverda to derive schema \dodoX from schema \taskyX by splitting a horizontal partition from \taskA{} with \sql{prio=1} and dropping the priority column.
Executing the script creates a new schema including the view \todos{} as well as delta code for propagating data changes.
When a new entry is inserted in \todos{}, this will automatically insert a corresponding task with priority $1$ to \taskA{} in \taskyX.
Equally, updates and deletions are propagated back to the \taskyX schema.
Pioneering work like PRIMA allows to create the version \dodoX as well, however the DEL is not bidirectional, hence write operations are only propagated from \taskyX to \dodoX but not vice versa.


For the next release \taskyyX, it is decided to normalize the table \taskA{} into \taskA{} and \authorA{}.
For a stepwise roll-out of \taskyyX, the old schema of \taskyX has to remain alive until all clients have been updated. Again, \inverda does the job.
When executing the \indel script as shown in Figure~\ref{fig:example}, \inverda creates the schema version \taskyyX and decomposes the table version \taskA{} to separate the tasks from their authors while creating a foreign key to maintain the dependency.
Additionally, the column \cAuthor{} is renamed to \cName.
\inverda generates delta code to make the \taskyyX schema immediately available.
Write operations to any of the three schema versions are propagated to all other versions.

\textbf{DBA: }
The initially materialized tables are the targets of create table SMOs.
All other table versions are implemented with the help of delta code.
The delta code introduces an overhead on read and write accesses to new schema versions.
The more SMOs are between schema versions, the more delta code is involved and the higher is the overhead.
In our example, the schema versions \taskyyX and \dodoX have delta code towards the physical table \taskA{}.
Some weeks after releasing \taskyyX the majority of the users has upgraded to the new version.
\taskyyX comes with its own phone app, so the schemas \taskyX and \dodoX\ are still accessed but merely by a minority of users.
It seems appropriate to migrate data physically to the table versions of the \taskyyX schema, now.
Traditionally, developers would write a migration script, which moves the data and implements new delta code.
All that accumulates to some hundred lines of code, which need to be tested intensively in order to prevent it from messing up our data.
With \inverda, the DBA writes a single line:
\sqll{\sqlk{MATERIALIZE} \sqlc{\taskyy}; }%
Upon this statement, \inverda transparently runs the physical data migration to schema \taskyyX, maintaining transaction guarantees, and updates the involved delta code of all schema versions.
There is no need to involve any developer. 
All schema versions stay available; read and write operations are merely propagated to a different set of physical tables, now.
Again, these features are facilitated by our bidirectional \indel; other approaches either have to materialize all schema versions and provide only forward propagation of data (PRIMA) or have to materialize the latest version and stop serving the old version (PRISM).
In sum, \inverda allows the user to continuously use all schema versions, the developer to continuously develop the applications, and the DBA to independently optimize the physical table schema.

\section{\inverdaHeading Architecture}\label{sec:architecture}

\newcommand{\dashrule}[1][black]{%
  \color{#1}\rule[\dimexpr.5ex-.2pt]{4pt}{.4pt}\xleaders\hbox{\rule{4pt}{0pt}\rule[\dimexpr.5ex-.2pt]{4pt}{.4pt}}\hfill\kern0pt \\
}
\begin{figure}
\fontsize{9}{9.4}\selectfont
\begin{tabular}{@{\hspace{0.6mm}}l@{\hspace{0.6mm}}}
    \toprule
    \sql{\sqlk{CREATE SCHEMA VERSION} $name_{new}$ [\sqlk{FROM} $name_{old}$]}\\
    \sql{\phantom{CREATE}  \sqlk{WITH} $SMO_1;\ldots\; SMO_n;$} \\
    \sql{\sqlk{DROP SCHEMA VERSION} $version_n;$}\vspace{-1mm} \\ \vspace{-1mm}
    \dashrule
    \sql{\sqlk{CREATE TABLE} \tbl{$R$}{$c_1$,\ldots,$c_n$}} \\
    \sql{\sqlk{DROP TABLE} $R$}\\
    \sql{\sqlk{RENAME TABLE} $R$ \sqlk{INTO} $R'$}\\
    \sql{\sqlk{RENAME COLUMN} $r$ \sqlk{IN} $R_i$ \sqlk{TO} $r'$} \\
    \sql{\sqlk{ADD COLUMN} $a$ \sqlk{AS} \tbl{$f$}{$r_1$,\ldots,$r_n$} \sqlk{INTO} $R_i$} \\
    \sql{\sqlk{DROP COLUMN} $r$ \sqlk{FROM} $R_i$ \sqlk{DEFAULT} \tbl{$f$}{$r_1$,\ldots,$r_n$}}\\
    \sql{\sqlk{DECOMPOSE TABLE} $R$ \sqlk{INTO} \tbl{$S$}{$s_1,\ldots,s_n$}}\\
    \sql{\phantom{DECOMPOSE }[, \tbl{$T$}{$t_1,\ldots,t_m$} \sqlk{ON} (\sqlk{PK}|\sqlk{FK} $fk$|$cond$)]}\\
    \sql{[\sqlk{OUTER}] \sqlk{JOIN TABLE} $R$, $S$ \sqlk{INTO} $T$ \sqlk{ON} (\sqlk{PK}|\sqlk{FK} $fk$|$cond$)}\\
    \sql{\sqlk{\PARTITION TABLE} $T$ \sqlk{INTO} $R$ \sqlk{WITH} \cR [, $S$ \sqlk{WITH} \cS]}\\
    \sql{\sqlk{MERGE TABLE} $R$ (\cR), $S$ (\cS) \sqlk{INTO} $T$} \\
    \bottomrule
\end{tabular}\\[-2mm]
\caption{Syntax of \indel SMOs.}
\label{fig:smos}
\end{figure}

\inverda simply builds upon existing relational DBMSes.
It adds schema evolution functionality and support for co-existing schema versions.
\inverda functionality is exposed to users via two interfaces:
(1) \indel (bidirectional database evolution language) and
(2) migration commands.

\textbf{\indel} provides a comprehensive set of bidirectional SMOs to create a new schema version either from scratch or as an evolution from a given schema version.
SMOs evolve \emph{source} tables to {target} tables.
Each table version is created by one \emph{incoming SMO} and evolved by arbitrarily many \emph{outgoing SMOs}.
Specifically, \indel SMOs allow to create or drop or rename tables and columns, (vertically) decompose or join tables, and (horizontally) split or merge tables (Syntax in Figure~\ref{fig:smos}, general semantics in~\cite{CarloA.Curino,HVBL15}, bidirectional semantics in Section~\ref{sec:singlesmo}).
We restrict the considered expressiveness of \indel to the relational algebra; the evolution of further artifacts like constraints~\cite{Curino2012} and functions is promising future work. 
\indel is the youngest child in an evolution of DELs: PRISM~\cite{CarloA.Curino} is a practically comprehensive DEL that couples schema and data evolution, CoDEL~\cite{HVBL15} extended PRISM to be relationally complete, and \indel extends CoDEL to be bidirectional.
As a prerequisite for co-existing schema versions, the unique feature of \indel SMOs is bidirectionality.
Essentially, the arguments of each \indel SMO gather enough information to facilitate full propagation of reads and writes between schema versions in both directions, forward propagation from the old to the new version as well as backward propagation from the new to the old version.
For instance, \sql{DROP COLUMN} requires a function $f(r_1,\ldots,r_n)$ that computes the value for the dropped column if a tuple, inserted in the new schema version, is propagated back to an old schema version.
Finally, \indel allows dropping unnecessary schema versions, which drops the schema version itself but maintains the data if still needed in other versions. 

\inverda's \textbf{migration commands} allow for changing the physical data representation.
By default, data is materialized in the source schema version.
Assuming a table is split in a schema evolution step, the data remains physically unsplit.
With a migration command the physical data representation of this SMO instance can be changed so that the data is also physically split. 
Within this work, we focus on non-redundant materialization which means the data is stored either on the source or the target side of the SMO but not on both.
Migration commands are very simple.
They either materialize a set of table versions or a complete schema version.
The latter is merely a convenience command allowing to materialize multiple table versions in one step.

In our prototypical implementation~\cite{Herrmann2016a}, \inverda creates the co-existing schema versions with views and triggers in a common relational DBMS.
\inverda interacts merely over common DDL and DML statements, data is stored in regular tables, and database applications use the DBMS's standard query engine.
For data accesses of database applications, only the generated views and triggers are used and no \inverda components are involved.
The employed triggers can lead to a cascaded execution, but in a controlled manner as there are no cycles in the version history.
Thanks to this architecture, \inverda easily utilizes existing DBMS components such as physical data storage, indexing, transaction handling, query processing, etc.\ without reinventing the wheel.


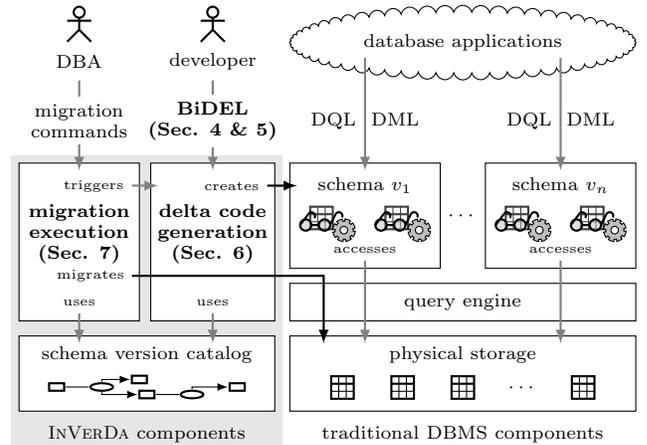
\begin{figure}
\centering
\begin{tikzpicture}[node font=\scriptsize,font=\scriptsize,
every node/.style={font=\scriptsize},
component/.style={every node,draw,outer sep=1mm,fill=white},
every edge/.style={draw=black!50, thick, -latex},
user/.pic={
    \coordinate (-crutch)       at (0,2mm);
    \coordinate (-south)        at (0,0);
    \coordinate (-left foot)    at (-2mm,0);
    \coordinate (-right foot)   at (2mm,0);
    \coordinate (-breast)       at (0,4mm);
    \coordinate (-left hand)    at (-2mm,3.5mm);
    \coordinate (-right hand)   at (2mm,3.5mm);
    \coordinate (-head)         at (0,6mm);
    \draw[thick] (-left hand) to (-right hand);
    \draw[thick] (-left foot) to (-crutch) to (-right foot);
    \draw[thick,-{Circle[open]}] (-crutch) to (-head);
  }
]

\newcommand{\contabicon}{
    \draw[thick] (0,0) rectangle (2mm,1mm);
}
\newcommand{\smoicon}{
    \draw[thick] (0,0) circle [x radius=1.5mm, y radius=0.5mm];
}

\node(cat)[component,minimum width=3.4cm,minimum height=1cm,align=center]{schema version catalog\\\\};

\node(contab1) [anchor=south,inner sep=0mm] at ($(cat.south)+(-12mm,3.5mm)$) {\tikz\contabicon;};
\node(smo1)    [anchor=south,inner sep=0mm] at ($(cat.south)+( -6mm,3.5mm)$) {\tikz\smoicon;};
\node(contab2) [anchor=south,inner sep=0mm] at ($(cat.south)+( -1mm,4.5mm)$) {\tikz\contabicon;};
\node(contab3) [anchor=south,inner sep=0mm] at ($(cat.south)+(  0mm,2.5mm)$) {\tikz\contabicon;};
\node(smo2)    [anchor=south,inner sep=0mm] at ($(cat.south)+(  6mm,2.5mm)$) {\tikz\smoicon;};
\node(contab4) [anchor=south,inner sep=0mm] at ($(cat.south)+( 12mm,3.5mm)$) {\tikz\contabicon;};
\draw[thin] (contab1) -- (smo1);
\draw[thin,-latex] (smo1) |- (contab2);
\draw[thin,-latex] (smo1) |- (contab3);
\draw[thin] (contab3) -- (smo2);
\draw[thin,-latex] (smo2) |- (contab4);

\node(migrexec)[component,minimum width=1.6cm,minimum height=2.1cm,align=center,anchor=south west,font=\bfseries] at (cat.north west) {migration\\execution\\(Sec. \ref{sec:migration})\\};
\node(dcodegen)[component,minimum width=1.6cm,minimum height=2.1cm,align=center,anchor=south east,font=\bfseries] at (cat.north east) {delta code\\generation\\(Sec. \ref{sec:generation})\\};
\draw[every edge] ($(migrexec.south)+(0,2mm)$)--($(cat.north-|migrexec.south)+(0,-2mm)$)
        node(migrexecusecat) [at start,font=\tiny,anchor=south] {uses};
\draw[every edge] ($(dcodegen.south)+(0,2mm)$)--($(cat.north-|dcodegen.south)+(0,-2mm)$)
        node(dcodegenusecat) [at start,font=\tiny,anchor=south] {uses};

\node(dba) [anchor=south,minimum height=5mm] at ($(migrexec.north)+(0,10mm)$) {DBA};
\pic(dbaicon) [anchor=south,scale=0.8] at ($(dba.north)+(0,0mm)$) {user};
\node(dev) [anchor=south,minimum height=5mm] at ($(dcodegen.north)+(0,10mm)$) {developer};
\pic(devicon) [anchor=south,scale=0.8] at ($(dev.north)+(0,0mm)$) {user};
\draw[every edge] ($(migrexec.north)+(0,10mm)$)--($(migrexec.north)+(0,-2mm)$) node[pos=0.45,fill=white,align=center] {migration\\commands};
\draw[every edge] ($(dcodegen.north)+(0,10mm)$)--($(dcodegen.north)+(0,-2mm)$) node[pos=0.45,fill=white,align=center,font=\bfseries] {\indel\\(Sec. \ref{sec:singlesmo} \& \ref{sec:bidirectionality})};

\node(schema1)[component,minimum width=2cm,minimum height=1.4cm,align=center,anchor=north west] at (dcodegen.north east) {schema $v_1$\\\\\\};
\node(scheman)[component,minimum width=2cm,minimum height=1.4cm,align=center,anchor=north west] at ($(schema1.north east)+(4mm,0)$) {schema $v_n$\\\\\\};
\node(schemai)[outer ysep=1mm,inner xsep=0mm,minimum height=1.4cm,align=center,anchor=center] at ($(schema1.east)!0.5!(scheman.west)$) {$\cdots$};

\node(schema1view1) [anchor=south] at ($(schema1.south)+(-5mm,3.5mm)$) {\tikz\viewtrigger;};
\node(schema1view2) [anchor=south] at ($(schema1.south)+( 5mm,3.5mm)$) {\tikz\viewtrigger;};
\node(schemanview1) [anchor=south] at ($(scheman.south)+(-5mm,3.5mm)$) {\tikz\viewtrigger;};
\node(schemanview2) [anchor=south] at ($(scheman.south)+( 5mm,3.5mm)$) {\tikz\viewtrigger;};
\draw[every edge, draw=black] ($(dcodegen.east|-schema1.west)+(-2mm,4mm)$)--($(schema1.west)+(2mm,4mm)$)
        node[at start,font=\tiny,anchor=east] {creates};

\draw[every edge] ($(migrexec.east|-schema1.west)+(-2mm,4mm)$)--($(dcodegen.west|-schema1.west)+(2mm,4mm)$)
        node[at start,font=\tiny,anchor=east] {triggers};

\node(apps) [cloud,draw,cloud puffs=60,aspect=4.8,minimum width=4.6cm,anchor=south] at ($(schemai.north)+(0,10mm)$) {database applications};
\draw[every edge] ($(schema1.north|-apps.south)+(0,3mm)$)--($(schema1.north)+(0,-2mm)$)
        node(dqldmllbl) [pos=0.56] {DQL\hspace{2mm}DML};
\draw[every edge] ($(scheman.north|-apps.south)+(0,3mm)$)--($(scheman.north)+(0,-2mm)$)
        node[pos=0.56] {DQL\hspace{2mm}DML};

\node(queryex)[component,minimum width=4.6cm,minimum height=0.5cm,align=center,anchor=north] at (schemai.south) {query engine};

\node(storage)[component,minimum width=4.6cm,minimum height=1cm,align=center,anchor=north] at (queryex.south) {physical storage\\\\};

\node(tab1) [anchor=south] at ($(storage.south)+( -8mm,1.5mm)$) {\tikz\tab;};
\node(tab2) [anchor=south] at ($(storage.south)+(-16mm,1.5mm)$) {\tikz\tab;};
\node(tab3) [anchor=south] at ($(storage.south)+(  0mm,1.5mm)$) {\tikz\tab;};
\node(tabi) [anchor=south] at ($(storage.south)+(  8mm,2.5mm)$) {$\cdots$};
\node(tabn) [anchor=south] at ($(storage.south)+( 16mm,1.5mm)$) {\tikz\tab;};

\node(migrexecmigratesstorage)[font=\tiny,anchor=east] at ($(migrexec.east|-queryex.north)+(-2mm,0mm)$) {migrates};
\coordinate (x_tmp) at ($(storage.west)!0.5!(schema1.south)$);
\draw[every edge, draw=black] (migrexecmigratesstorage.east)-|($(x_tmp|-storage.north)+(0mm,-2mm)$);

\draw[every edge] ($(schema1.south)+(0,2mm)$)--($(storage.north-|schema1.south)+(0,-2mm)$)
        node[at start,font=\tiny,anchor=south] {accesses};
\draw[every edge] ($(scheman.south)+(0,2mm)$)--($(storage.north-|scheman.south)+(0,-2mm)$)
        node[at start,font=\tiny,anchor=south] {accesses};

\node(invcomplbl) [anchor=north] at (cat.south) {\inverda components};
        node[midway,below,yshift=-1mm] {traditional DBMS components};
\node(tradcomplbl) [anchor=north] at (storage.south) {traditional DBMS components};

\begin{pgfonlayer}{background}
    \node(getsrc)[fill=gray!20,inner sep=0mm,fit=(migrexec)(dcodegen)(cat)(invcomplbl)] {};
\end{pgfonlayer}


\end{tikzpicture}\\[-2mm]
\caption{\inverda integration into DBMS.}
\label{fig:architecture}
\vspace{-3mm}
\end{figure}

\begin{figure}
\centering
\begin{tikzpicture}[node font=\scriptsize,font=\scriptsize,
every node/.style={font=\scriptsize},
component/.style={every node,draw,outer sep=1mm},
catedge/.style={draw, darkgray, thin, -latex},
every pin/.append style={font=\tiny,text=gray},
every pin edge/.style={gray,thin,shorten <=1pt}]
]

\newcommand{\tcontab}[5]{
    \node(#1) [anchor=south,inner sep=1mm, draw #5] at ($(#3 mm,#4 mm)$) {\elt{#2}};
}

\newcommand{\tmatcontab}[5]{
    \node(#1) [anchor=south,inner sep=1mm, draw, fill=black!20 #5] at ($(#3 mm,#4 mm)$) {\elt{#2}};
}
\newcommand{\tsmo}[5]{
    \node(#1) [anchor=south,inner sep=1mm, draw, rounded corners=4pt,font=\tiny\ttfamily #5] at ($(#3 mm,#4 mm)$) {#2};
}

\tmatcontab{c1}{Task}{4}{0}{,pin=150:Table Version in Phy. Storage}
\tsmo{s1}{\PARTITION}{-6}{-5}{}
\tcontab{c2}{Todo}{-13}{-10}{,pin=120:Table Version}
\tsmo{s2}{DROP COLUMN}{-22}{-15}{}
\tcontab{c3}{Todo}{-35}{-15}{}
\tsmo{s3}{DECOMPOSE}{15}{-5}{,pin=40:SMO instance}
\tcontab{c4}{Task}{30}{-10}{}
\tcontab{c5}{Author}{5}{-10}{}
\tsmo{s4}{RENAME COLUMN}{15}{-15}{}
\tcontab{c6}{Author}{30}{-15}{}

\draw[catedge] (c1) -| (s1);
\draw[catedge] (s1) |- (c2);
\draw[catedge] (c2) |- (s2);
\draw[catedge] (s2) -- (c3);

\draw[catedge] (c1) -| (s3);
\draw[catedge] (s3) |- (c4);
\draw[catedge] (s3) |- (c5);
\draw[catedge] (c5) |- (s4);
\draw[catedge] (s4) -- (c6);

\node(c1lbl) [anchor=south west,inner sep=0mm,yshift=1mm] at (c1.north west) {\taskyX};
\node(c1box) [inner sep=1mm,draw,dashed,fit=(c1)(c1lbl),pin=15:Schema Version] {};
\node(c3lbl) [anchor=south west,inner sep=0mm,yshift=1mm] at (c3.north west) {\dodoX};
\node(c3box) [inner sep=1mm,draw,dashed,fit=(c3)(c3lbl)] {};
\node(c4c6lbl) [anchor=south west,inner sep=0mm,yshift=1mm] at (c4.north-|c6.west) {\taskyyX};
\node(c4c6box) [inner sep=1mm,draw,dashed,fit=(c4)(c6)(c4c6lbl)] {};

%

\end{tikzpicture}\\[-2mm]
\caption{Example of schema version catalog.}
\label{fig:catalog}
\end{figure}
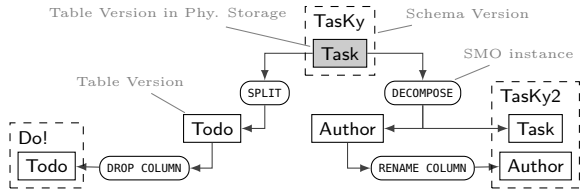

Figure~\ref{fig:architecture} outlines the principle components of an \inverda-equipped DBMS.
As can be seen, \inverda adds three components to the DBMS:
(1) the \emph{delta code generation} creates views and triggers to expose schema versions based on the current physical data representation.
Delta code generation is either triggered by a developer issuing \indel commands to create a new schema version or by the DBA issuing migration commands to change the physical data representation.
The delta code consists of standard commands of the DBMS's query engine.
(2) the \emph{migration execution} orchestrates the actual migration of data from one physical representation to another and the adaptation of the delta code.
The data migration is done with the help of query engine capabilities.
(3) the \emph{schema version catalog} maintains the \emph{genealogy of schema versions}:
It is the history of all schema versions including all table versions as well as the SMO instances and their materialization state.
Figure~\ref{fig:catalog} shows the schema version catalog for our \taskyX example with the initial materialization. 

When developers execute \indel scripts, the respective SMO instances and table versions are registered in the schema version catalog.
The schema version catalog maintains references to tables in the physical storage that hold the payload data and to auxiliary tables that hold otherwise lost information of the not necessarily information preserving SMOs.
The materialization states of the SMOs, which can be changed by the DBA through migration commands, determine which data tables and auxiliary tables are physically present and which are not.
\inverda uses the schema version catalog to generate delta code for new schema versions or for changed physical table schemas.
Data accesses of applications are processed by the generated delta code within the DBMS's query engine.
When a developer drops an old schema version that is not used any longer, the schema version is removed from the catalog.
However, the respective SMOs are only removed from the catalog in case they are no longer part of an evolution that connects two remaining schema versions.


The schema version catalog is the central knowledge base for all schema versions and the evolution between them.
To this end, the catalog stores the genealogy of schema versions by means of a directed acyclic hypergraph $(T,E)$.
Each vertex $t \in T$ represents a table version.
Each hyperedge $e \in E$ represents one SMO instance, i.e., one table evolution step.
An SMO instance $e=(S,T)$ evolves a set of source table versions $S$ into a set of target table versions $T$.
Additionally, the schema version catalog stores for every SMO instance the SMO type (split, merge, etc.), the parameter set, and its state of materialization.
Each schema version is a subset of all table versions in the system.
Schema versions share a table version if the table evolves in-between them.
At evolution time, \inverda uses the catalog to generate delta code that makes all schema versions accessible.
At query time, the generated delta code itself is executed by the existing DBMS's query engine---outside \inverda components.

\section{\indelHeading - Bidirectional SMO{\normalsize \textbf{s}}}\label{sec:singlesmo}

\indel{}'s unique feature is the bidirectional semantics of its SMOs, which is the basis for \inverda's co-existing schema versions.
We highlight the design principles behind \indel SMOs and formally validate their bidirectionality.
All \indel SMOs follow the same design principles.
Without loss of generality, the \sql{\PARTITION} SMO is used as a representative example in this section to explain the concepts.
The remaining SMOs are introduced in Appendix~\ref{apx:SMOs}.

Figure~\ref{fig:smomapping} illustrates the principle structure of a single SMO instance resulting from the sample statement
\sqll{\sqlk{SPLIT TABLE} $T$ \sqlk{INTO} $R$ \sqlk{WITH} \cR, $S$ \sqlk{WITH} \cS}%
which horizontally splits a source table $T$ into two target tables $R$ and $S$ based on conditions \cR and \cS.
Assuming both schemas are materialized, reads and writes on both schema versions can simply be delegated to the corresponding data tables $T_D$, $R_D$, and $S_D$, respectively.
However, \inverda materializes data non-redundantly on one side of the SMO instance, only.
If the data is physically stored on the source side of an SMO instance, the SMO instance is called \emph{virtualized}; with data stored on the target side it is called \emph{materialized}.
In any case, reads and writes on the unmaterialized side are mapped to the materialized side.

The semantics of each SMO is defined by two functions \getTarget and \getSource which describe precisely the mapping from the source side to the target side and vice versa, respectively.
Assuming the target side of \sql{\PARTITION} is materialized, all reads on $T$ are mapped by \getSource to reads on $R_D$ and $S_D$; and writes on $T$ are mapped by \getTarget to writes on $R_D$ and $S_D$.
While the payload data of $R$, $S$, and $T$ is stored in the physical tables $R_D$, $S_D$, and $T_D$, the tables $R^-$, $S^+$, $S^-$, $R^*$, $S^*$, and $T'$ are auxiliary tables for the \sql{\PARTITION} SMO to prevent information loss.
Note that the semantics of SMOs is complete, if reads and writes on both source and target schema work correctly regardless on which of both sides the data is physically stored.
This basically means that each schema version acts like a full-fledged database schema; however it does not enforce that data written in any version is also fully readable in other versions.
In fact, \indel ensures this for all SMOs except of those that create redundancy---in these cases the developer specifies a preferred replica beforehand.

Obviously, there are different ways of defining \getTarget and \getSource; in this paper, we propose one way that systematically covers all potential inconsistencies and is bidirectional.
We aim at a non-redundant materialization, which also includes that the auxiliary tables merely store the minimal set of required auxiliary information.
Starting from the basic semantics of the SMO---e.g. the splitting of a table---we incrementally detect inconsistencies that contradict the bidirectional semantics and introduce respective auxiliary tables.
The proposed rule sets can serve as a blueprint, since they clearly outline which information needs to be stored to achieve bidirectionality.

To define \getTarget and \getSource, we use Datalog---a compact and solid formalism that facilitates both a formal evaluation of bidirectionality and easy delta code generation.
Precisely, we use Datalog rule templates instantiated with the parameters of an SMO instance.
For brevity of presentation, we use some extensions to the standard Datalog syntax:
For variables, small letters represent single attributes and capital letters lists of attributes.
For equality predicates on attribute lists, both lists need to have the same length and same content, i.e.~for $A=(a_1,\ldots,a_n)$ and $B=(b_1,\ldots,b_m)$, $A=B$ holds if $n=m\wedge a_1=b_1 \wedge\ldots\wedge a_n=b_n$.
All tables have an attribute \pk, an \inverda-managed identifier to uniquely identify tuples across versions.
Additionally, \pk ensures that the multiset semantics of a relational database fits with the set semantics of Datalog, as the unique key \pk prevents equal tuples in one relation.
For a table $T$ we assume $T(\pk,\_)$ and $\neg T(\pk,\_)$ to be safe predicates since any table can be projected to its key.

For the exemplary \sql{\PARTITION}, let's assume this SMO instance is materialized, i.e.\ data is stored on the target side, and let's consider the \getTarget mapping function first.
\sql{\PARTITION} horizontally splits a table $T$ from the source schema into two tables $R$ and $S$ in the target schema based on conditions \cR and \cS:
\begin{align}
R(\pk,A)&\leftarrow T(\pk,A),\cR (A)\label{rule:rSimple}\\
S(\pk,A)&\leftarrow T(\pk,A),\cS (A)\label{rule:sSimple}
\end{align}
The conditions \cR and \cS can be arbitrarily set by the user so that Rule~\ref{rule:rSimple} and Rule~\ref{rule:sSimple} are insufficient wrt.\ the desired bidirectional semantics, since the source table $T$ may contain tuples neither captured by \cR nor by \cS.
In order to avoid inconsistencies and make the SMO bidirectional, such tuples are stored on the target side in the auxiliary table $T'$:
\begin{align}
T'(\pk,A)&\leftarrow T(\pk,A),\neg \cR (A),\neg \cS (A)\label{rule:T'Simple}
\end{align}

\begin{figure}
\centering
\begin{tikzpicture}[node font=\scriptsize,font=\scriptsize,
every node/.style={font=\scriptsize},
opedge/.style={draw, thick, -latex},
catedge/.style={draw, darkgray, thin, -latex},
tab/.style={draw,minimum height=4mm,inner sep=0mm},
contab/.style={tab,minimum width=6mm,fill=white},
dattab/.style={tab,minimum width=5mm,fill=white},
auxtab/.style={tab,minimum width=5mm,fill=white},
cattab/.style={tab,minimum width=3mm,fill=white,font=\tiny},
tabbox/.style={draw,dashed,fill=gray!20},
getop/.style={fill=white},
getopedge/.style={opedge,latex-latex}
]

\newcommand{\viewsandtriggers}[1]{
    \node[scale=0.7] at ($(#1.north west)+(0.5mm,-0.4mm)$) {\tikz\view;};
    \node[scale=0.9] at ($(#1.south east)+(0, 0.5mm)$) {\tikz\gear;};
}

\node(contabT) [contab] {$\,T$};
\viewsandtriggers{contabT}
\node(datatabT) [dattab,anchor=north] at ($(contabT.south)+(0,-8mm)$) {$T_D$};
\node(auxtabsSp) [auxtab,anchor=north] at ($(datatabT.south)+(0,-1mm)$) {$S^+$};
\node(auxtabsRp) [auxtab,anchor=east] at ($(auxtabsSp.west)+(-1mm,0)$) {$R^-$};
\node(auxtabsSm) [auxtab,anchor=west] at ($(auxtabsSp.east)+( 1mm,0)$) {$S^-$};
\node(auxtabsRs) [auxtab,anchor=north east] at ($(auxtabsSp.south)+(-.5mm,-1mm)$) {$R^*$};
\node(auxtabsSs) [auxtab,anchor=north west] at ($(auxtabsSp.south)+(.5mm,-1mm)$) {$S^*$};

\node(contabR) [contab,anchor=south west] at ($(contabT.east)+(38mm,1mm)$) {$R$};
\viewsandtriggers{contabR}
\node(contabS) [contab,anchor=north] at ($(contabR.south)+(0,-2mm)$) {$S$};
\viewsandtriggers{contabS}
\node(datatabR) [dattab,anchor=east,xshift=-0.5mm] at (datatabT.west-|contabS.south) {$R_D$};
\node(datatabS) [dattab,anchor=west,xshift= 0.5mm] at (datatabT.west-|contabS.south) {$S_D$};
\node(auxtabsTp) [auxtab,anchor=center] at ($(auxtabsSp.south-|contabS.south)+(0mm,-.5mm)$) {$T'$};

\node(schemaSrc) [anchor=south] at ($(contabT.north)+(0,5mm)$) {source schema $v_i$};
\node(schemaTrg) [anchor=base] at (schemaSrc.base-|contabR.north) {target schema $v_{i+1}$};

\begin{pgfonlayer}{background}
    \node (tbox) [tabbox,fit=(contabT),minimum width=20mm,minimum height=13mm] {};
    \node (phytbox) [tabbox,fit=(datatabT)(auxtabsRp)(auxtabsSm)(auxtabsRs)(auxtabsSs),minimum width=20mm,minimum height=13mm] {};
    \node (rsbox) [tabbox,fit=(contabR)(contabS),minimum width=14mm,minimum height=13mm] {};
    \node (phyrsbox) [tabbox,fit=(datatabR)(datatabS)(auxtabsTp)(auxtabsSs.south-|auxtabsTp),minimum width=14mm,minimum height=11mm] {};
\end{pgfonlayer}

\draw[opedge,-latex] ($(phytbox.north)+(-6mm,-3mm)$)--($(tbox.north)+(-6mm,-3mm)$)
    node[anchor=south,inner sep=0.5mm] {read};
\draw[opedge,latex-] ($(phytbox.north)+( 6mm,-3mm)$)--($(tbox.north)+( 6mm,-3mm)$)
    node[anchor=south,inner sep=0.5mm] {write};

\coordinate (x_smo) at ($(tbox.east)!0.5!(rsbox.west)$);
\node(partsmo) [draw,darkgray,thin,rounded corners=4pt,inner sep=1mm,font=\tiny\ttfamily] at (contabT.east-|x_smo) {\PARTITION};
\begin{pgfonlayer}{background}
    \draw[catedge] (contabT.east) -- (partsmo.west);
    \draw[catedge] (partsmo.north) |- (contabR.west);
    \draw[catedge] (partsmo.south) |- (contabS.west);
\end{pgfonlayer}
\node(catlbl) [anchor=north,darkgray,inner sep=1mm,font=\tiny] at (tbox.north-|x_smo) {schema catalog};

\coordinate (x_cat) at ($(tbox.west)+(-.5mm,0)$);
\node(contabslbl) [align=center,anchor=east,font=\tiny] at (x_cat|-contabT.center) {views\\\&\\triggers};
\node(dattabsblb) [align=center,anchor=center,font=\tiny] at (contabslbl|-datatabT) {data\\tables};
\node(auxtabsblb) [align=center,anchor=center,font=\tiny] at (contabslbl|-auxtabsTp) {auxiliary\\tables};

\coordinate (x_level) at ($(contabslbl.west)+(-0mm,0)$);
\node(catlbl) [rotate=90,anchor=south,align=center,font=\tiny] at (x_level|-contabslbl.west) {schemas\\versions};
\coordinate (y_storagelbl) at ($(dattabsblb)!0.5!(auxtabsblb)$);
\node(storlbl)[rotate=90,anchor=south,align=center,font=\tiny] at (x_level|-y_storagelbl) {physical\\storage};

\coordinate (y_phyBorder) at ($(tbox.south)!0.5!(phytbox.north)$);
\coordinate (x1_phyBorder) at ($(catlbl.north)+(1mm,0)$);
\coordinate (x2_phyBorder) at ($(schemaTrg.east)+(-1mm,0)$);

\begin{pgfonlayer}{background}
    \draw[dotted] (x1_phyBorder|-y_phyBorder) -- (x2_phyBorder|-y_phyBorder);
\end{pgfonlayer}

\draw[getopedge] ($(phytbox.north east)+(0,-2mm)$)
    .. controls ($(x_smo|-phytbox.north east)+(0,-2mm)$) and ($(x_smo|-tbox.south east)+(0,2mm)$) ..
    ($(rsbox.south west)+(0,2mm)$);
\begin{pgfonlayer}{background}
    \node(gettrg)[getop,anchor=base west,yshift=-0.5mm] at (tbox.east|-y_phyBorder) {\getSource};
\end{pgfonlayer}

\draw[getopedge] ($(phyrsbox.north west)+(0,-2mm)$)
    .. controls ($(x_smo|-phyrsbox.north west)+(0,-2mm)$) and ($(x_smo|-rsbox.south west)+(0,2mm)$) ..
    ($(tbox.south east)+(0,2mm)$);
\begin{pgfonlayer}{background}
    \node(getsrc)[getop,anchor=base east,yshift=-0.5mm] at (rsbox.west|-y_phyBorder) {\getTarget};
\end{pgfonlayer}

\node(remark)[align=center,darkgray,inner sep=1mm,font=\tiny] at (partsmo.south|-auxtabsSs.west) {alternative\\materializations};
\draw[thin,gray,shorten >=1mm,shorten <=-3mm] (remark)--(phytbox);
\draw[thin,gray,shorten >=1mm,shorten <=-3mm] (remark)--(phyrsbox);

\end{tikzpicture}\\[-2mm]
\caption{Mapping functions of single \sql{\PARTITION} SMO.}
\label{fig:smomapping}
\vspace{-1mm}
\end{figure}
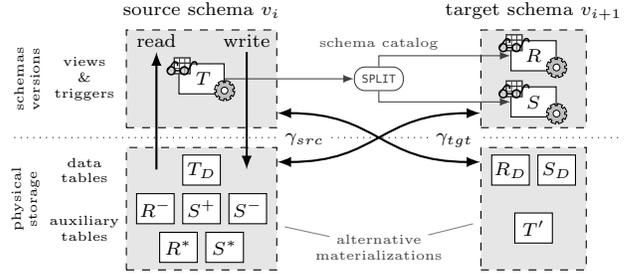

Let's now consider the \getSource mapping function for reconstructing $T$ while the target side is still considered to be materialized.
Reconstructing $T$ from the target side is essentially a union of $R$, $S$, and $T'$.
Nevertheless, $c_R$ and $c_S$ are not necessarily disjoint. 
One source tuple may occur as two equal but independent instances in $R$ and $S$.
We call such two instances \emph{twins}.
Twins can be updated independently resulting in \emph{separated twins}, i.e.\ two tuples---one in $R$ and one in $S$---with equal key \pk but different value for the other attributes.
To resolve this ambiguity and make the SMO bidirectional, we consider the first twin in $R$ to be the primus inter pares and define \getSource of \sql{\PARTITION} to propagate back all tuples in $R$ as well as those tuples in $S$ not contained in $R$:
\begin{align}
T(\pk,A)&\leftarrow R(\pk,A)\label{rule:tFromR}\\
T(\pk,A)&\leftarrow S(\pk,A), \neg R(\pk,\_)\label{rule:tFromS}\\
T(\pk,A)&\leftarrow T'(\pk,A)\label{rule:tFromTprime}
\end{align}
The Rules~\ref{rule:rSimple}--\ref{rule:tFromTprime} define sufficient semantics for \sql{\PARTITION} as long as the target side is materialized.

Let's now assume the SMO instance is virtualized, i.e.\ data is stored on the source side, and let's keep considering the \getSource mapping function.
Again, $R$ and $S$ can contain separated twins---unequal tuples with equal key \pk.
According to Rule~\ref{rule:tFromS}, $T$ stores only the separated twin from $R$.
To avoid losing the other twin in $S$, it is stored in the auxiliary table $S^+$:
\begin{align}
S^+(\pk,A)&\leftarrow S(\pk,A), R(\pk,A'),A\neq A'\label{rule:S+FromS}
\end{align}
Accordingly, \getTarget has to reconstruct the separated twin in $S$ from $S^+$ instead of $T$ (concerns Rule~\ref{rule:sSimple}).
Twins can also be deleted independently resulting in a \emph{lost twin}.
Given the data is materialized on the source side, a lost twin would be directly recreated from its other twin via $T$.
To avoid this information gain and keep lost twins lost, \getSource keeps the keys of lost twins from $R$ and $S$ in auxiliary tables $R^-$ and $S^-$:
\begin{align}
R^-(\pk) &\leftarrow S(\pk,A),\neg R(\pk,\_),\cR (A)\\
S^-(\pk) &\leftarrow R(\pk,A),\neg S(\pk,\_),\cS (A)
\end{align}
Accordingly, \getTarget has to exclude lost twins stored in $R^-$ from $R$ (concerns Rule~\ref{rule:rSimple}) and those in $S^-$ from $S$ (concerns Rule~\ref{rule:sSimple}).
Twins result from data changes issued to the target schema containing $R$ and $S$ which can also lead to tuples that do not meet the conditions \cR resp. \cS.
In order to ensure that the reconstruction of such tuples is possible from a materialized table $T$, auxiliary tables $R^*$ and $S^*$ are employed for identifying those tuples using their identifiers (concerns Rules~\ref{rule:rSimple} and~\ref{rule:sSimple}).
\begin{align}
S^*(\pk) &\leftarrow S(\pk,A),\neg \cS(A)\\
R^*(\pk) &\leftarrow R(\pk,A),\neg \cR(A)
\end{align}
The full rule sets of \getTarget respectively \getSource are now bidirectional and
defined as follows:
\begin{align}
\mathbf{\getTarget:}&\nonumber\\
R(\pk,A)&\leftarrow T(\pk,A),\cR (A), \neg R^-(\pk)\label{rule:getrg:begin}\\
R(\pk,A)&\leftarrow T(\pk,A), R^*(\pk)\label{rule:getrg:rs}\\
S(\pk,A)&\leftarrow T(\pk,A),\cS (A), \neg S^-(\pk), \neg S^+(\pk,\_)\\
S(\pk,A)&\leftarrow S^+(\pk,A)\\
S(\pk,A)&\leftarrow T(\pk,A), S^*(\pk), \neg S^+(\pk,\_)\label{rule:getrg:ss}\\
T'(\pk,A)&\leftarrow T(\pk,A),\neg \cR (A),\neg \cS (A), \neg  R^*(\pk), \neg  S^*(\pk)\label{rule:getrg:end}\\
\mathbf{\getSource:}&\nonumber\\
T(\pk,A)&\leftarrow R(\pk,A)\label{rule:tFromRFinal}\\
T(\pk,A)&\leftarrow S(\pk,A), \neg R(\pk,\_)\label{rule:tFromSFinal}\\
T(\pk,A)&\leftarrow T'(\pk,A)\label{rule:tFromTprimeFinal}\\
R^-(\pk) &\leftarrow S(\pk,A),\neg R(\pk,\_),\cR (A)\\
R^*(\pk) &\leftarrow R(\pk,A),\neg \cR(A)\\
S^+(\pk,A)&\leftarrow S(\pk,A), R(\pk,A'),A\neq A'\\
S^-(\pk) &\leftarrow R(\pk,A),\neg S(\pk,\_),\cS (A)\\
S^*(\pk) &\leftarrow S(\pk,A),\neg \cS(A)\label{rule:getsrc:end}
\end{align}
The semantics of all other \indel SMOs are defined in a similar way, see Appendix~\ref{apx:SMOs}.
This precise definition of \indel's SMOs, is the basis for the formal validation of their bidirectionality.

\newpage
\section{Formal evaluation of \indelHeadings\newline bidirectionality}\label{sec:bidirectionality}
\indel's SMOs are bidirectional, because, no matter whether the data is (1) materialized on the source side (SMO is virtualized) or (2) materialized on the target side (SMO is materialized), both sides behave like a full-fledged single-schema database.
To formally evaluate this claim, we consider the two cases (1) and (2) independently.
Let's start with case (1); the data is materialized on the source side.
For a correct target-side propagation, the data \dt at the target side has to be mapped by \getSource to the data tables and auxiliary tables at the source side (write) and mapped back by \getTarget to the data tables on the target side (read) without any loss or gain visible in the data tables at the target side.
Similar conditions have already been defined for symmetric relational lenses~\cite{Hofmann2011}---given data at the target side, storing it at the source side, and mapping it back to target should return the identical data at the target side.
For the second case (2) it is vice versa.
Formally, an SMO has bidirectional semantics if the following holds:
\begin{align}
\dt&=\getTarget^{data}(\getSource(\dt))\label{eq:dt}\\
\ds&=\getSource^{data}(\getTarget(\ds))\label{eq:ds}
\end{align}
Data tables that are visible to the user need to match these bidirectionality conditions.
As indicated by the index $\gamma^{data}$, we project away potentially created auxiliary tables; however, they are always empty except for SMOs that calculate new values: e.g. adding a column requires to store the calculated values when data is stored at the source side to ensure repeatable reads.
The bidirectionality conditions are shown by applying and simplifying the Datalog rule sets that define the mappings \getSource and \getTarget.
We label the original relations to distinguish them from the resulting relation, apply \getSource and \getTarget in the order according to Condition~\ref{eq:dt} or~\ref{eq:ds}, and compare the outcome to the original relation.
It has to be identical.
As neither the rules for a single SMO nor the version genealogy have cycles, there is no recursion at all, which simplifies evaluating the combined Datalog rules.

In the following, we introduce some basic notion about Datalog rules as basis for the formal evaluation.
A Datalog rule is a clause of the form $H\leftarrow L_1,\ldots,L_n$ with $n \geq 1$ where $H$ is an atom denoting the rule's head, and $L_1,\ldots,L_n$ are literals, i.e. positive or negative atoms, representing its body.
For a given rule $r$, we use \rhead{r} to denote its head $H$ and \rbody{r} to denote its set of body literals $L_1,\ldots,L_n$.
In the mapping rules defining \getSource and \getTarget, every \rhead{r} is of the form $q^r(\pk,Y)$ where $q^r$ is the derived predicate, \pk{} is the \inverda-managed identifier, and $Y$ is a potentially empty list of variables.
Further, we use \rpred{r} to refer to the predicate symbol of \rhead{r}.
For a set of rules \ruleset{R}, $\ruleset{R}^q$ is defined as $\set{r \mid r\in\ruleset{R} \wedge \rpred{r}=q}$.
For a body literal $L$, we use \rpred{L} to refer to the predicate symbol of $L$ and \rvars{L} to denote the set of variables occurring in $L$.
In the mapping rules, every literal $L\in\rbody{r}$ is of the form either $q^r_i(\pk,Y^r_i,X^r_i)$ or $c^r(Y^r_i,X^r_i)$, where $Y^r_i\subset Y$ are the variables occurring in $L$ and \rhead{r} and $X^r_i$ are the variables occurring in $L$ but not in \rhead{r}.
Generally, we use capital letters to denote multiple variables.
For a set of literals $\mathcal{K}$, \rvars{\mathcal{K}} denotes $\bigcup_{L \in \mathcal{K}}\rvars{L}$.
The following lemmas are used for simplifying a given rule set \ruleset{R} into a rule set \ruleset{R'} such that \ruleset{R'} derives the same facts as \ruleset{R}.

\begin{lemma}[Deduction]\label{lemma:deducation}
Let $L\equiv q^r(\pk,Y)$ be a literal in the body of a rule $r$.
For a rule $s\in\ruleset{R}^{\rpred{L}}$ let $\renm{s,L}$ be rule $s$ with all variables occurring in the head of $s$ at positions of $Y$ variables in $L$ be renamed to match the corresponding $Y$ variable and all other variables be renamed to anything not in \rvars{\rbody{r}}.
If $L$ is 
\begin{enumerate}[leftmargin=*] 
\item a positive literal, $s$ can be applied to $r$ to get rule set $r(s)$ of the form $\set{\rhead{r} \leftarrow \rbody{r}\setminus\set{L} \cup \rbody{\renm{s,L}}}$.
\item a negative literal, $s$ can be applied to $r$ to get rule set $r(s)=$ $\set{\rhead{r} \leftarrow \rbody{r}\setminus\set{L} \cup t(K) \mid K \in \rbody{\renm{s,L}})}$\\ with either $t(K)=\set{\neg q^s_i(\pk,Y^s_i,\_)}$ if $K\equiv q^s_i(\pk,Y^s_i,X^s_i)$ or 
$t(K)=\{q^s_j(\pk,Y^s_j,X^s_j) \mid q^s_j(\pk,Y^s_i,X^s_j)\in\rbody{\renm{s,L}} \wedge X^s_j\cap X^s_i\neq\emptyset\}\cup\set{c^r(Y^s_i,X^s_i)}$ if $K\equiv c^r(Y^s_i,X^s_i)$.\footnote{Correctness can be shown with help of first order logic.}
\end{enumerate}
For a given $p$, let $r$ be every rule in \ruleset{R} having a literal $L\equiv p(X,Y)$ in its body.
Accordingly, \ruleset{R} can be simplified by replacing all rules $r$ and all $s\in\ruleset{R}^p$ with all $r(s)$ applications to $\ruleset{R} \setminus (\set{r}\cup\ruleset{R}^p) \cup (\bigcup_{s\in\ruleset{R}^{\rpred{L}}}r(s))$.
\end{lemma}

\begin{lemma}[Empty Predicate]\label{lemma:empty}
Let $r\in \ruleset{R}$ be a rule, $L$ be a literal in the body $L\in\rbody{r}$ and the relation \rpred{L} is known to be empty.
If $L$ is a positive literal, $r$ can be removed from \ruleset{R}.
If $L$ is a negative literal, $r$ can be simplified to $\rhead{r}\leftarrow\rbody{r}\setminus\set{L}$.
\end{lemma}

\begin{lemma}[Tautology]\label{lemma:tautology}
Let $r,s\in\ruleset{R}$ be rules and $L$ and $K$ be literals in the bodies of $r$ and $s$, respectively, where $r$ and $s$ are identical except for $L$ and $K$, i.e. $\rhead{r}=\rhead{s}$ and $\rbody{r}\setminus\set{L}=\rbody{s}\setminus\set{K}$, or can be renamed to be so.
If $K\equiv\neg L$, $r$ can be simplified to $\rhead{r}\leftarrow\rbody{r}\setminus\set{L}$ and $s$ can be removed from \ruleset{R}.
\end{lemma}

\begin{lemma}[Contradiction]\label{lemma:unsatisfiability}
Let $r\in \ruleset{R}$ be a rule and $L$ and $K$ be literals in its body $L,K\in\rbody{r}$.
If $K\equiv\neg L$, $r$ can be removed from \ruleset{R}.
\end{lemma}



\begin{lemma}[Unique Key]\label{lemma:p}
Let $r\in \ruleset{R}$ be a rule and $q(\pk,X)$ and $q(\pk,Y)$ be literals in its body.
Since, by definition, \pk is a unique identifier, $r$ can be modified to $\rhead{r}\leftarrow\rbody{r}\cup\set{X=Y}$.
\end{lemma}

In this paper, we use these lemmas to show bidirectionality for the materialized \sql{\PARTITION} SMO in detail.
Hence, Equation~\ref{eq:ds} needs to be satisfied.
Writing data $T_D$ from source to target-side results in the mapping $\getTarget(\Tdata)$.
With target-side materialization all source-side auxiliary tables are empty.
Thus, $\getTarget(\Tdata)$ can be simplified with Lemma~\ref{lemma:empty}:
\begin{align}
R(\pk,A)&\leftarrow \Tdata(\pk,A),\cR (A)\\
S(\pk,A)&\leftarrow \Tdata(\pk,A),\cS (A)\\
T'(\pk,A)&\leftarrow \Tdata(\pk,A),\neg \cR (A),\neg \cS (A)
\end{align}
Reading the source-side data back from $R$, $S$, and $T'$ to $T$ adds the rule set \getSource (Rule~\ref{rule:tFromRFinal}--\ref{rule:getsrc:end}) to the mapping.
Using Lemma~\ref{lemma:deducation}, the mapping $\getSource(\getTarget(\Tdata))$ simplifies to:
\begin{align}
T(\pk,A)\leftarrow& \Tdata(\pk,A),\cR(A) \label{align:pr1}\\
T(\pk,A)\leftarrow& \Tdata(\pk,A),\cS(A),\neg\Tdata(\pk,A) \label{align:partition_Source_T_empty}\\
T(\pk,A)\leftarrow& \Tdata(\pk,A),\cS(A),\neg\cR(A) \label{align:ps}\\
T(\pk,A)\leftarrow& \Tdata(\pk,A),\neg\cS(A),\neg\cR(A) \label{align:ns}\\
R^-(\pk) \leftarrow& \Tdata(\pk,A),\cS(A),\neg\Tdata(\pk,A),\cR(A)\\
R^-(\pk) \leftarrow& \Tdata(\pk,A),\cS(A),\neg\cR (A),\cR (A) \label{align:rminus}\\
R^*(\pk) \leftarrow& \Tdata(\pk,A),\cR(A),\neg\cR(A) \label{align:rstar}
\end{align}
\begin{align}
S^+(\pk,A)\leftarrow& \Tdata(\pk,A),\cS(A),\Tdata(\pk,A'),\cR(A'), A\!\neq\!A'\label{align:splus}\\
S^-(\pk) \leftarrow& \Tdata(\pk,A),\cR(A),\neg\Tdata(\pk,A),\cS(A) \label{align:sminus}\\
S^-(\pk) \leftarrow& \Tdata(\pk,A),\cR(A),\neg\cS (A),\cS (A)\\
S^*(\pk) \leftarrow& \Tdata(\pk,A),\cS(A),\neg\cS(A)\label{align:sstar}
\end{align}
With Lemma~\ref{lemma:unsatisfiability}, we omit Rule~\ref{align:partition_Source_T_empty} as it contains a contradiction.
With Lemma~\ref{lemma:tautology}, we reduce Rules~\ref{align:ps} and~\ref{align:ns} to Rule~\ref{align:nr} by removing the literal $c_S(A)$.
The resulting rules for $T$
\begin{align}
T(\pk,A)&\leftarrow \Tdata(\pk,A),\cR (A)\label{align:pr2}\\
T(\pk,A)&\leftarrow \Tdata(\pk,A),\neg \cR (A) \label{align:nr}
\end{align}
can be simplified again with Lemma~\ref{lemma:tautology} to
\begin{align}
T(\pk,A)&\leftarrow \Tdata(\pk,A)\quad. \label{partition_Source_T_logic_final}
\end{align}
For Rule~\ref{align:splus}, Lemma~\ref{lemma:p} implies $A=A'$, so this rule can be removed based on Lemma~\ref{lemma:unsatisfiability}.
Likewise, the Rules \ref{align:rminus}--\ref{align:sstar} have contradicting literals on \Tdata, \cR, and \cS respectively so that Lemma~\ref{lemma:unsatisfiability} applies here as well.
The result clearly shows that data $\Tdata$ in \ds is mapped by $\getSource(\getTarget(\ds))$ to the target side and back to \ds without any information loss or gain:
\begin{align}
\mathbf{\getSource(\getTarget(\ds)):}\;\;
 &T(\pk,A)\leftarrow \Tdata(\pk,A) \qed
\end{align}
So, $\ds=\getSource(\getTarget(\ds))$ holds.
Remember that the auxiliary tables only exist on the materialized side of the SMO (target in this case).
Hence, it is correct that there are no rules left producing data for the source-side auxiliary.
The same can be done for Equation~\ref{eq:dt} as well (Appendix~A).
As expected, the simplification of $\getTarget(\getSource(\dt))$ results in
\begin{align}
\mathbf{\getTarget(\getSource(\dt)):}\;\;
&R(\pk,A)\leftarrow \Rdata(\pk,A)\\
&S(\pk,A)\leftarrow \Sdata(\pk,A)
\end{align}

This formal evaluation works for the remaining \indel SMOs, as well (Appendix~\ref{apx:SMOs}).
\indel's SMOs ensure that given data at any schema version $V_{n}$ that is propagated and stored at a direct predecessor $V_{n-1}$ or direct successor schema version $V_{n+1}$ can always be read completely and correctly in $V_n$.
To our best knowledge, we are the first to design a set of powerful SMOs and validate their bidirectionality according to the criteria of symmetric relational lenses.

\textbf{Write operations: }
Bidirectionality also holds after write operations: When updating a not-materialized schema version, this update is propagated to the materialized schema in a way that it is correctly reflected when reading the updated data again.
Given a materialized SMO, we apply a write operation $\Delta_{src}(\ds)$ to given data on the source side.
$\Delta_{src}(\ds)$ can both insert and update and delete data.
Initially, we store \ds at the target side using $\dt=\getTarget(\ds)$.
To write at the source side, we have to temporarily map back the data to the source with $\getSource^{data}(\dt)$, apply the write $\Delta_{src}$, and map the updated data back to target with $\dt'=\getTarget(\Delta_{src}(\getSource^{data}(\dt)))$.
Reading the data from the updated target $\getSource^{data}(\dt')$ has to be equal to applying the write operation $\Delta_{src}(\ds)$ directly on the source side.
\begin{align}
\Delta_{src}(\ds)&=\getSource^{data}(\getTarget(\Delta_{src}(\getSource^{data}(\getTarget(\ds)))))\label{cond:1}
\end{align}
We have already shown that $D=\getSource^{data}(\getTarget(D))$ holds for any data $D$ at the target side, so that Equation~\ref{cond:1} reduces to $\Delta_{src}(\ds) = \Delta_{src}(\ds)$.
Hence writes are correctly propagated through the SMOs.
The same holds vice versa for writing at the target-side of virtualized SMOs:
\begin{align}
\Delta_{tgt}(\dt)&=\getTarget^{data}(\getSource(\Delta_{tgt}(\getTarget^{data}(\getSource(\dt)))))
\end{align}


\textbf{Chains of SMOs: }
Further, the bidirectionality of \indel SMOs also holds for chains of SMOs: $smo_1,\ldots smo_n$, where $\getX{i, src/tgt}$ is the respective mapping of $smo_i$.
Analogous to symmetric relational lenses~\cite{Hofmann2011}, there are no side-effects between multiple \indel SMOs.
So, \indel's bidirectionality is also guaranteed along chains of SMOs:
\begin{align}
\dt&=\gamma_{n,tgt}^{data}(\ldots\gamma_{1,tgt}(\gamma_{1,src}(\ldots\gamma_{n,src}(\dt))))\label{eq:dtChain}\\
\ds&=\gamma_{1,src}^{data}(\ldots\gamma_{n,src}(\gamma_{n,tgt}(\ldots\gamma_{1,tgt}(\ds))))\label{eq:dsChain}
\end{align}
This bidirectionality ensures logical data independence, since any schema version can now be read and written without information loss or gain, no matter where the data is actually stored.
The auxiliary tables keep the otherwise lost information and we have formally validated their feasibility. 
With the formal guarantee of bidirectionality---also along chains of SMOs and for write operations---we have laid a solid formal foundation for \inverda's delta code generation.

\section{Delta Code Generation}\label{sec:generation}


\begin{figure}
\centering
\definecolor{darkgreen}{rgb}{0,.5,0}
\begin{tikzpicture}[node font=\scriptsize,font=\scriptsize,
every node/.style={font=\scriptsize},
tab/.style={draw,minimum height=4mm,minimum width=9mm,inner sep=0mm,fill=white},
tabs/.style={inner sep=0},
cattab/.style={tab,minimum width=3mm,fill=white,font=\tiny},
getopedge/.style={draw,thick},
smo/.style={draw,darkgray,thin,ellipse,fill=white,font=\tiny\ttfamily,minimum width=10mm,minimum height=3mm,inner sep=.1mm},
smoedge/.style={darkgray,thin},
bar/.style={draw,darkgray,fill=darkgray,inner sep=0mm},
barv/.style={bar,minimum width=.5mm,minimum height=5mm},
barh/.style={bar,minimum width=8mm,minimum height=.5mm},
matstate/.style={font=\sffamily\bfseries,gray},
delta/.style={draw,regular polygon,regular polygon sides=3,fill=white,inner sep=0.6mm,yshift=-0.6mm}
]

\newcommand{\tables}[1]{
    \node [tab] at ( 1mm, -1mm) {};
    \node [tab] at (.5mm,-.5mm) {};
    \node [tab] at ( 0mm,  0mm) {#1};
}
\newcommand{\viewsandtriggers}[1]{
    \node[scale=0.7] at ($(#1.north west)+(1.3mm,-.4mm)$) {\tikz\view;};
    \node[scale=0.9] at ($(#1.south east)+(-1mm, 1mm)$) {\tikz\gear;};
}
\newcommand{\phytab}[1]{
    \node[scale=0.7] at ($(#1.south east)+(-1mm, 1mm)$) {\tikz\tab;};
}

\coordinate (x_contabs0) at (0,0);
\coordinate (x_contabs0_b) at ($(x_contabs0)+(2mm,0)$);

\node(contabs1) [tabs,anchor=west] at ($(x_contabs0.east)+(17mm,0mm)$) {\tikz\tables{$\,T_{i-1}$};};
\viewsandtriggers{contabs1}

\node(contabs2) [tabs,anchor=west] at ($(contabs1.east)+(11mm,0mm)$) {\tikz\tables{$\,T_{i}$};};
\viewsandtriggers{contabs2}
\node(datatabs2) [tabs,anchor=north] at ($(contabs2.south)+(0,-9mm)$) {\tikz\tables{$D_{i}$};};
\phytab{datatabs2}
\node(auxtabs2a) [tabs,anchor=north] at ($(datatabs2.south)+(0,-1mm)$) {\tikz\tables{$A_{i}$};};
\phytab{auxtabs2a}
\node(auxtabs2b) [tabs,anchor=north] at ($(auxtabs2a.south)+(0,-1mm)$) {\tikz\tables{$A_{i+1}$};};
\phytab{auxtabs2b}

\node(contabs3) [tabs,anchor=west] at ($(contabs2.east)+(11mm,0mm)$) {\tikz\tables{$\,T_{i+1}$};};
\viewsandtriggers{contabs3}
\node(auxtabs3a) [tabs,anchor=center] at (auxtabs2a.west-|contabs3.south) {\tikz\tables{$A_{i+2}$};};
\phytab{auxtabs3a}

\coordinate (x_contabs4) at ($(contabs3.east)+(17mm,0mm)$);
\coordinate (x_contabs4_b) at ($(x_contabs4)+(-2mm,0)$);

\node(contabs1smo0con) [barv,anchor=east] at ($(contabs1.west)+(-1mm,0)$) {};
\node(contabs2smo1con) [barv,anchor=east] at ($(contabs2.west)+(-1mm,0)$) {};
\node(contabs2smo2con) [barv,anchor=west] at ($(contabs2.east)+( 1mm,0)$) {};
\node(contabs3smo3con) [barv,anchor=west] at ($(contabs3.east)+( 1mm,0)$) {};
\node(auxtabs2asmo1con) [barv,anchor=east] at ($(auxtabs2a.west)+(-1mm,0)$) {};
\node(auxtabs2bsmo2con) [barv,anchor=west] at ($(auxtabs2b.east)+( 1mm,0)$) {};
\node(auxtabs3asmo1con) [barv,anchor=west] at ($(auxtabs3a.east)+( 1mm,0)$) {};

\node(contabs1smo1con) [barh,anchor=north] at ($(contabs1.south)+(0,-1mm)$) {};
\node(contabs2datatabs2con) [barh,anchor=north] at ($(contabs2.south)+(0,-1mm)$) {};
\node(datatabs2contabs2con) [barh,anchor=south] at ($(datatabs2.north)+(0,1mm)$) {};
\node(contabs3smo2con) [barh,anchor=north] at ($(contabs3.south)+(0,-1mm)$) {};

\coordinate (y_phyBorder) at ($(contabs2.south)!0.5!(datatabs2.north)$);
\coordinate (x1_phyBorder) at ($(x_contabs0_b)+(0mm,0)$);
\coordinate (x2_phyBorder) at ($(x_contabs4_b)+(0mm,0)$);

\begin{pgfonlayer}{background}
    \draw[dotted] (x1_phyBorder|-y_phyBorder) -- (x2_phyBorder|-y_phyBorder);
\end{pgfonlayer}

\node(case2) [anchor=south,text=blue] at ($(contabs1.north)+(0,7mm)$) {\underline{Case 2}};
\node(case1) [anchor=base,text=darkgreen] at (case2.base-|contabs2.north) {\underline{Case 1}};
\node(case3) [anchor=base,text=red] at (case2.base-|contabs3.north) {\underline{Case 3}};


\coordinate (y_smo)  at ($(contabs1.north)+(0,3mm)$);
\coordinate (x_smo0) at ($(x_contabs0)!0.5!(contabs1.west)$);
\coordinate (x_smo1) at ($(contabs1.east)!0.5!(contabs2.west)$);
\coordinate (x_smo2) at ($(contabs2.east)!0.5!(contabs3.west)$);
\coordinate (x_smo3) at ($(contabs3.east)!0.5!(x_contabs4)$);

\node(smo0) [smo,anchor=center] at (y_smo-|x_smo0) {$\text{SMO}_{i-1}$};
\node(smo1) [smo,anchor=center] at (smo0.west-|x_smo1) {$\text{SMO}_{i}$};
\node(smo2) [smo,anchor=center] at (smo1.west-|x_smo2) {$\text{SMO}_{i+1}$};
\node(smo3) [smo,anchor=center] at (smo2.west-|x_smo3) {$\text{SMO}_{i+2}$};

\draw[smoedge,dotted] (x_contabs0_b|-smo0)--(smo0);
\draw[smoedge,-latex] (smo0)-|(contabs1.105);
\draw[smoedge,-latex] (contabs1.75)|-(smo1)-|(contabs2.105);
\draw[smoedge,-latex] (contabs2.75)|-(smo2)-|(contabs3.105);
\draw[smoedge] (contabs3.75)|-(smo3);
\draw[smoedge,dotted] (smo3)--(x_contabs4_b|-smo3);

\node[matstate,anchor=south] at (smo0.north) {materialized};
\node[matstate,anchor=south] at (smo1.north) {materialized};
\node[matstate,anchor=south] at (smo2.north) {virtualized};
\node[matstate,anchor=south] at (smo3.north) {virtualized};

\coordinate (y_op) at ($(contabs2.south)!0.6!(datatabs2.north)$);

\draw[getopedge,latex-] (contabs1smo0con.west)-|(x_smo0|-y_op);
\draw[getopedge] (x_smo0|-y_op)--(smo0.200|-y_op);
\draw[getopedge,dotted] (smo0.200|-y_op)--(x_contabs0_b|-y_op);
\draw[getopedge,blue,latex-] (contabs2smo1con.west)-|(x_smo1|-y_op);
\draw[getopedge,blue,latex-] (auxtabs2asmo1con.west)-|(x_smo1|-y_op);
\draw[getopedge,blue] (x_smo1|-y_op)-|(contabs1smo1con.south) node[delta] {};
\draw[getopedge,red,latex-] (contabs2smo2con.east)-|(x_smo2|-y_op);
\draw[getopedge,red,latex-] (auxtabs2bsmo2con.east)-|(x_smo2|-y_op);
\draw[getopedge,red] (x_smo2|-y_op)-|(contabs3smo2con.south) node[delta] {};
\draw[getopedge,latex-] (contabs3smo3con.east)-|(x_smo3|-y_op);
\draw[getopedge,latex-] (auxtabs3asmo1con.east)-|(x_smo3|-y_op);
\draw[getopedge] (x_smo3|-y_op)--(smo3.340|-y_op);
\draw[getopedge,dotted] (smo3.340|-y_op)-|(x_contabs4_b|-y_op);
\draw[getopedge,darkgreen,latex-] (datatabs2contabs2con.north)--(contabs2datatabs2con.south) node[delta] {};

\coordinate (x_level) at ($(x_contabs0_b)+(2mm,0)$);
\node(catlbl) [rotate=90,anchor=west,font=\tiny,inner ysep=0mm] at ($(x_level|-y_phyBorder)+(0,1mm)$) {schemas};
\node(storlbl)[rotate=90,anchor=east,font=\tiny,inner ysep=0mm] at ($(x_level|-y_phyBorder)+(0,-1mm)$) {physical storage};

\end{tikzpicture}\\[-3mm]
\caption{Three different cases in delta code generation.}
\label{fig:smochaining}
\vspace{-2mm}
\end{figure}
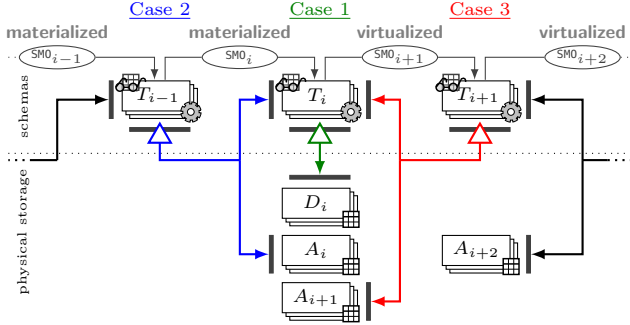

To make a schema version available, \inverda translates the \getSource and \getTarget mapping functions into delta code---specifically views and triggers.
Views implement delta code for reading; triggers implement delta code for writing.
In a schema versions genealogy, a single table version is the target of one SMO instance and the source for a number of SMO instances.
The delta code for a specific table version depends on the materialization state of the table's adjacent SMOs.

To determine the right rule sets for delta code generation, consider the schema genealogy in Figure~\ref{fig:smochaining}.
Table version $T_i$ is materialized, hence the two subsequent SMO instances, $i-1$ and $i$ store their data at the target side (materialized), while the two subsequent SMO instances, $i+1$ and $i+2$ are set to source-side materialization (virtualized).
Without loss of generality, three cases for delta code generation can be distinguished, depending on the direction a specific table version needs to go for to reach the materialized data.
\begin{description}[leftmargin=6mm] 
\item [Case 1 -- local:] The incoming SMO is materialized and all outgoing SMOs are virtualized.
The data of $T_i$ is stored in the data table $D_i$ and is directly accessible.
\item [Case 2 -- forwards:] The incoming SMO and one outgoing SMO are materialized.
The data of $T_{i-1}$ is stored in newer versions along the schema genealogy, so data access is propagated with \getSource (read) and \getTarget (write) of $\sql{SMO}_{i}$.
\item[Case 3 -- backwards:] The incoming SMO and all outgoing SMOs are virtualized.
The data of $T_{i+1}$ is stored in older versions along the schema genealogy, so data access is propagated with \getTarget (read) and \getSource (write) of $\sql{SMO}_{i+1}$.
\end{description}

\begin{figure}
\centering
\includegraphics[width=\columnwidth, page=9, trim=0mm 103mm 95mm 0mm, clip=true]{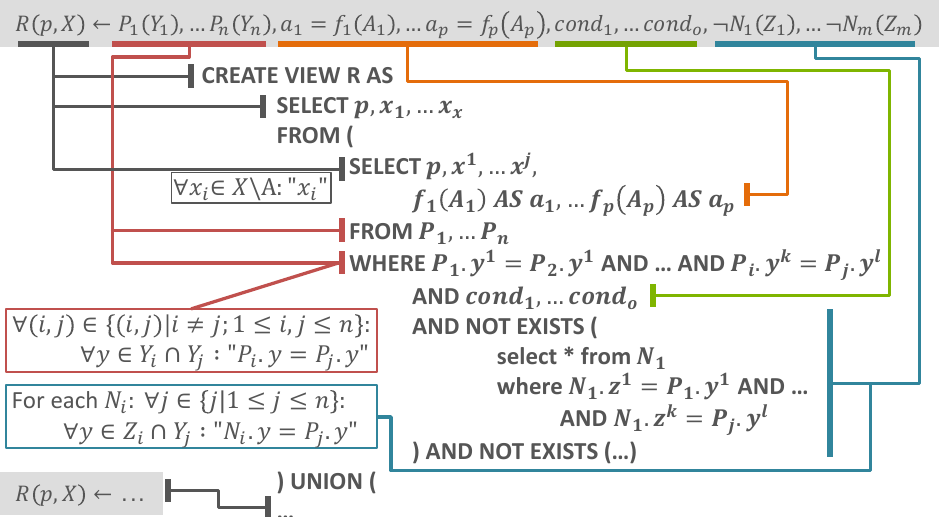}\\[-2mm]
\caption{SQL generation from Datalog rules.}
\label{fig:generation}
\end{figure}

In Case~1, delta code generation is trivial.
In Case~2 and ~3, \inverda essentially translates the Datalog rules defining the relevant mapping functions into view and trigger definitions.
Figure~\ref{fig:generation} illustrates the general pattern of the translation of Datalog rules to a view definition.
As a single table can be derived by multiple rules, e.g. Rule~\ref{rule:tFromRFinal}--\ref{rule:tFromTprimeFinal}, a view is a union of subqueries each representing one of the rules.
For each subquery, \inverda lists all attributes of the rule head in the select clause.
Within a nested subselect these attributes are either projected from the respective table version or derived by a function. 
All positive literals referring to other table versions or auxiliary tables are listed in the from clause.
Further, \inverda adds for all attributes occurring in multiple positive literals respective join conditions to the where clause.
Finally, conditions, such as $c_S(X)$, and negative literals, which \inverda adds as a \texttt{NOT EXISTS(<subselect for the literal>)} condition, complete the where-clause.

For writing, \inverda generates three triggers on each table version: for inserts, deletes, and updates.
To not recompute all data of the materialized side after each write operation at the not-materialized side of an SMO, \inverda adopts an update propagation technique for Datalog rules~\cite{Behrend2015} that results in minimal write operations.
For instance, an insert operation $\Delta_T^+(\pk,A)$ on the table version $T$ propagated back to the source side of a materialized \sql{\PARTITION} SMO results in the following update rules:
{\small
\begin{align}
\Delta^+_R(\pk,A)\leftarrow \Delta^+_T(\pk,A),&\mathtt{new}\;\cR(A),\mathtt{old}\;\neg R(\pk,A)\\
\Delta^+_S(\pk,A)\leftarrow \Delta^+_T(\pk,A),&\mathtt{new}\;\cS(A),\mathtt{old}\;\neg S(\pk,A)\\
\Delta^+_{T'}(\pk,A)\leftarrow \Delta^+_T(\pk,A),&\mathtt{new}\;\neg\cR(A),\neg\cS(A),\mathtt{old}\; \neg T'(\pk,A)
\end{align}
}%
The propagation can handle multiple write operations at the same time and distinguishes between \op{old} and \op{new} data, which represents the state before and after applying other write operations.
The derived update rules match the intuitive expectations:
The inserted tuple is propagated to $R$ or to $S$ or to $T'$ given it satisfies either $\cR$ or $\cS$ or none of them.
The additional conditions on the $\mathtt{old}$ literals ensure minimality by checking whether the tuple already exists. 
To generate trigger code from the update rule, \inverda applies essentially the same algorithm as for view generation.

Writes performed by a trigger on a table version further trigger the propagation along the schema genealogy to the other table versions as long as the respective update rules deduce write operations, i.e. as long as some data is physically stored with a table version either in the data table or in auxiliary tables.
With the generated delta code, \inverda propagates writes on any schema version to every other co-existing schema version in a schema genealogy.

\clearpage
\section{Migration Procedure}\label{sec:migration}
The materialization states of all SMO instances in a schema genealogy form the materialization schema.
The materialization schema determines the physical table schema, i.e. which table versions are directly stored in physical storage.
For the \taskyX example---Figure~\ref{fig:example}---this entails five different possible materialization schemas $M$, each implying a different physical table schema $P$ as shown in Table~\ref{tab:materializationSchemas}.

\begin{table}
\centering
\scalebox{0.9}{
\begin{tabular}{ll}
\toprule
$M$ &  $P$ \\
\midrule
$\emptyset$ & \set{\taskR{0}} \\ 
\set{\sql{\PARTITION}} & \set{\taskR{0}} \\
\set{\sql{\PARTITION},\sql{DROP COLUMN}} & \set{\todoR{1}} \\ 
\set{\sql{DECOMPOSE}} & \set{\taskR{1}, \authorR{0}} \\ 
\set{\sql{DECOMPOSE},\sql{RENAME}} & \set{\taskR{1}, \authorR{1}} \\ 
\bottomrule
\end{tabular}
}
\caption{Possible materialization schemas and the corresponding physical table schema in the \taskyX example.}
\label{tab:materializationSchemas}
\end{table}

The materialization schema has huge impact on the performance of a given workload.
Workload changes such as increased usage of newer schema versions demand adaptations of the materialization schema.
\inverda facilitates such an adaptation with a foolproof migration command.
The migration command allows moving data non-redundantly along the schema genealogy to those table versions where the given workload causes the least overhead for propagating reads and writes.
Initially, all SMOs except of the create table SMOs are virtualized, i.e. only initially created table versions are in the physical table schema.
A new materialization schema is derived from a given valid materialization schema by changing the materialization state of selected SMO instances.
Formally, two conditions must hold for a materialization schema to be valid.
For an SMO $s$, we denote the source table versions as \srctabs{s}.
For each table version $t$ we denote the incoming SMO with \insmo{t} and the set of outgoing SMOs with \outsmo{t}.
A materialization is valid iff:
{
\begin{align}
&\forall s\!\in\!M\ \forall t\!\in\!\srctabs{s}\ \brk{\insmo{t} \in M}\\
&\forall s\!\in\!M\ \forall t\!\in\!\srctabs{s}\ \nexists o\!\in\!\brk{\outsmo{t}\!\setminus\!\set{s}}\ \brk{o \in M}
\end{align}
}%
The first condition ensures that all source table versions are in the materialization schema.
The second condition ensures that no source table version is already taken by another materialized SMO.

In the migration command, the DBA lists the table versions that should be materialized. For instance:
\sqll{\sqlk{MATERIALIZE} \sqlc{\taskyy.task}, \sqlc{\taskyy.author}; }%
\inverda determines with the schema version catalog the corresponding materialization schema and checks, whether it is valid according to the conditions above.
If valid, \inverda automatically creates the new physical tables including auxiliary tables in the physical storage, migrates the data, regenerates all necessary delta code, and deletes all old physical tables.
The actual data migration relies on the same SQL generation routines as used for view generation.
From a user perspective, all schema versions still behave the same after a migration. 
However, any data access is now propagated to the new physical table schema resulting in a better performance for schema versions that are evolution-wise close to this new physical schema.
This migration is triggered by one single line of code, so adaptation to the current workload becomes a comfortable thing to do for database administrators.

\section{Evaluation}\label{sec:evaluation}
\inverda brings huge advantages for software systems by decoupling the different goals of different stakeholders.
Users can continuously access all schema versions, while developers can focus on the actual continuous implementation of the software without caring about former versions.
Above all, the DBA can change the physical table schema of the database to optimize the overall performance without restricting the availability of the co-existing schema versions or invalidating the developers' code.
In Section~\ref{sec:simplicity}, we show how \inverda reduces the length and complexity of the code to be written by the developer and thereby yields more robust and maintainable solutions.
\inverda automatically generates the delta code based on the discussed Datalog rules.
In Section~\ref{sec:overhead}, we measure the overhead of accessing data through \inverda's delta code compared to a handwritten SQL implementation of co-existing schema versions and show that it is reasonable.
In Section~\ref{sec:independence}, we show that the possibility to easily adapt the physical table schema to a changed workload outweighs the small overhead of \inverda's automatically generated delta code.
Materializing the data according to the most accessed version speeds up the data access significantly.

\begin{table}
\centering\scriptsize
\begin{tabular}{lrrr}
\toprule
\textbf{\inverda}&Initially&Evolution&Migration \\
\midrule
Lines of Code&1&3&1\\
Statements&1&3&1\\
Characters&54&152&19\\
\midrule
\\[-1mm]
\textbf{SQL (Ratio)}&Initially&Evolution&Migration \\
\midrule
Lines of Code&1 ($\times$1.00) & 359 ($\times$119.67) & 182 ($\times$182.00)\\
Statements&1 ($\times$1.00) & 148 ($\times$49.33) & 79 ($\times$79.00)\\
Characters&54 ($\times$1.00) & 9477 ($\times$62.35) & 4229 ($\times$222.58)\\
\bottomrule
\end{tabular}
\vspace{-2mm}
\caption{Ratio between SQL and \inverda delta code.}
\label{tab:simplicity}
\end{table}

\textbf{Setup:} For the measurements, we use three different data sets to gather a holistic idea of \inverda's characteristics.
We use (1) our \taskyX example as a middle-sized and comprehensive scenario, (2) \num{171} schema versions of Wikimedia~\cite{CarloA.Curino} as a long real-world scenario, and (3) short synthetic scenarios for all possible combinations of two SMOs.
We measure single thread performance of a PostgreSQL 9.4 database with co-existing schema versions on a Core i7 machine with 2,4GHz and 8GB memory.

\subsection{Simplicity and Robustness}\label{sec:simplicity}
Most importantly, we show that \inverda unburdens developers by rendering the expensive and error-prone task of manually writing delta code unnecessary. 
We show this using both the \taskyX example and Wikimedia.

\textbf{\taskyX:} We implement the evolution from \taskyX to \taskyyX with handwritten and hand-optimized SQL and compare this code to the equivalent \indel statements.
We manually implemented (1) creating the initial \taskyX schema, (2) creating the additional schema version \taskyyX with the respective views and triggers, and (3) migrating the physical table schema to \taskyyX and adapting all existing delta code.
%
This handwritten SQL code is much longer and much more complex than achieving the same goal with \indel.
Table~\ref{tab:simplicity} shows the lines of code (LOC) required with SQL and \indel, respectively, as well as the ratio between these values.
As there is no general coding style for SQL, LOC is a rather vague measure.
We also include the number of statements and the number of characters (consecutive white-space characters counted as one) as more objective measures to get a clear picture. 
Obviously, creating the initial schema is equally complex for both approaches.
However, evolving to the new schema version \taskyyX and migrating the data accordingly requires \num{359} and \num{182} lines of SQL code respectively, while we can express the same with \num{3} and \num{1} lines with \indel.
Moreover, the SQL code is also more complex, as indicated by the average number of character per statement.
While \indel is working exclusively on the visible schema versions, with handwritten SQL developers also have to manage auxiliary tables, triggers, etc.

The automated delta code generation does not only eliminate the error-prone and expensive manual implementation, but it is also reasonably fast.
Creating the initial \taskyX took \num{154} ms on our test system.
The evolution to \taskyyX, which includes two SMOs, requires \num{230} ms for both the generation and execution of the evolution script.
The same took \SI{177}{\milli\second} for \dodoX.
Please note that the complexity of generating and executing evolution scripts depends linearly on the number of SMOs $N$ and the number of untouched table versions $M$.
The complexity is $O(N+M)$, since we generate the delta code for each SMO locally and exclusively work on the neighboring table versions.
This principle protects from additional complexity in longer chains of SMOs.
The same holds for the complexity of executing migration scripts.
It is $O(N)$ since \inverda merely moves the data and updates the delta code for the materialized SMOs stepwise.

\begin{table}
\centering
\scalebox{0.7}{
\begin{tabular}{l l}
\toprule
\textbf{SMO}&occurrences \\
\midrule
\sql{CREATE TABLE}&42\\
\sql{DROP TABLE}&10\\
\sql{RENAME TABLE}&1\\
\sql{ADD COLUMN}&95\\
\sql{DROP COLUMN}&21\\
\bottomrule
\end{tabular}
}
\hspace{1cm}
\scalebox{0.7}{
\begin{tabular}{l l}
\toprule
\textbf{SMO}&occurrences \\
\midrule
\sql{RENAME COLUMN}&36\\
\sql{JOIN}&0\\
\sql{DECOMPOSE}&4\\
\sql{MERGE}&2\\
\sql{\PARTITION}&0\\
\bottomrule
\end{tabular}
}\vspace{-1mm}
\caption{Used SMOs in Wikimedia database evolution.}
\label{tab:wikimediaSMOs}
\end{table}

\begin{figure}[b]
\includegraphics[width=\linewidth, page=1, trim=0mm 10mm 0mm 20mm, clip=true]{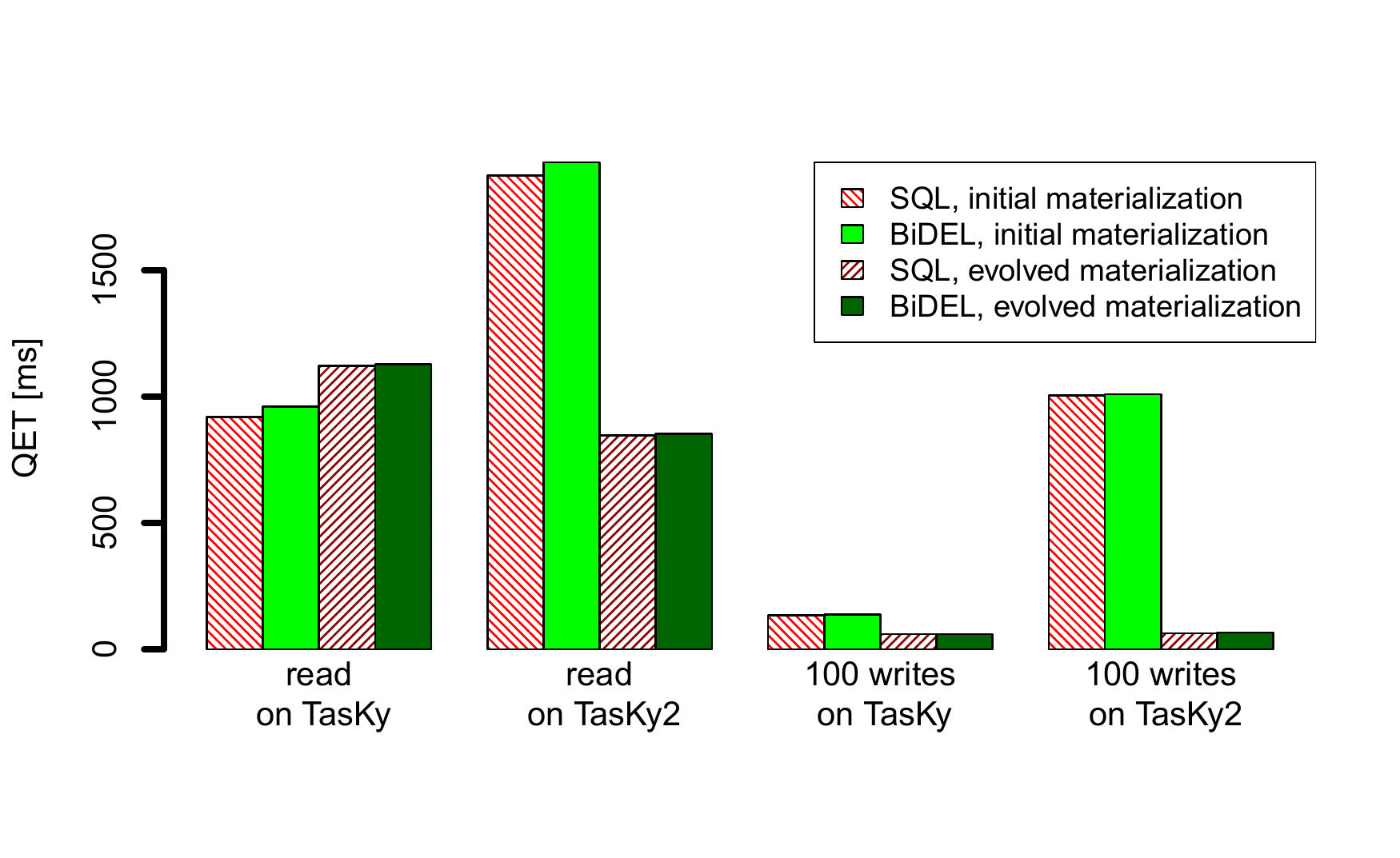}\vspace{-5mm}
\caption{Overhead of generated code.}
\label{fig:taskyOverhead}
\end{figure}

\textbf{Wikimedia:} Even long evolutions can be easily modeled with \indel.
To show this, we implement \num{171} schema versions of the Wikimedia~\cite{CarloA.Curino}, so data that is written in any of these schema versions, is also visible in all \num{170} other schema versions.
\indel proved to be capable of providing the database schema in each version exactly according to the benchmark and migrating the data accordingly. 
In Table~\ref{tab:wikimediaSMOs}, we summarize how often each SMO has been used in the 211 SMOs long evolution.
Even though simple SMOs, like adding and removing tables/columns, are clearly dominating---probably due to the restricted database evolution support of current DBMSes---there are more complex evolutions including the other SMOs as well.
Hence, there is a need for more sophisticated database evolution support and \indel shows to be feasible.

\begin{figure}
\includegraphics[width=\linewidth, page=1, trim=0mm 0mm 0mm 20mm, clip=true]{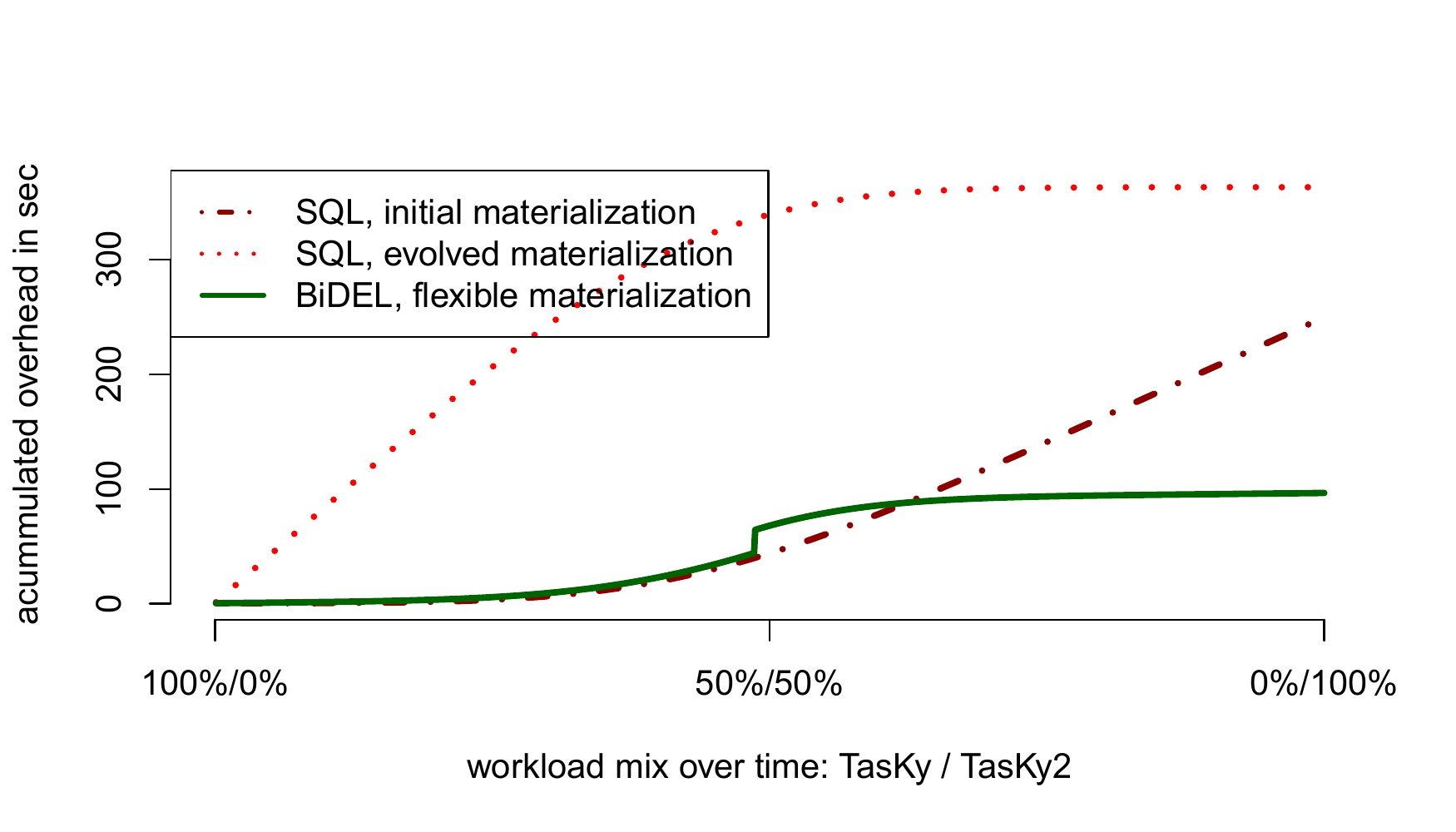}\vspace{-5mm}
\caption{Flexible materialization.}
\label{fig:scenario2}
\end{figure}

\begin{figure}
\includegraphics[width=\linewidth, page=1, trim=0mm 0mm 0mm 20mm, clip=true]{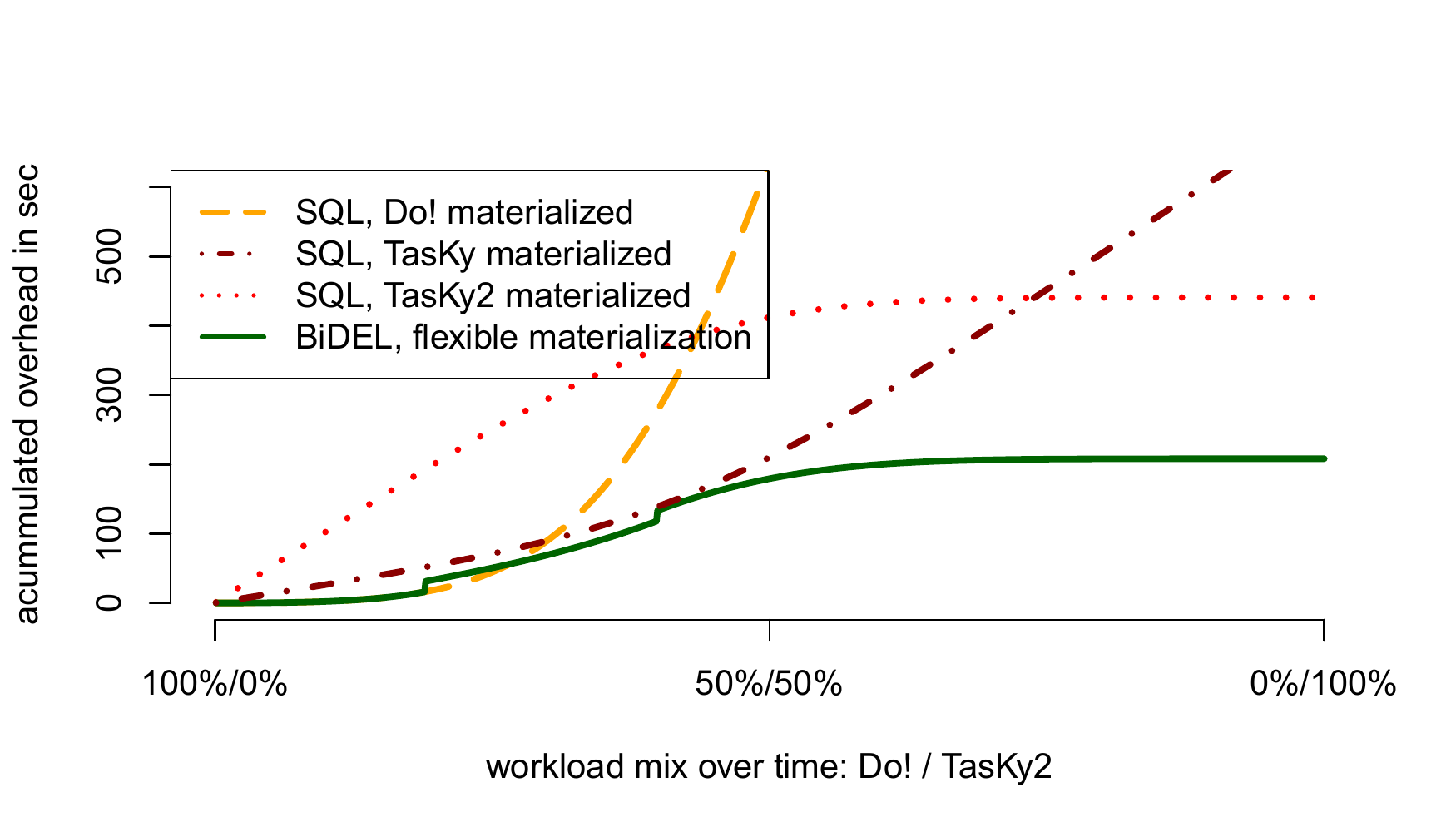}\vspace{-5mm}
\caption{Flexible materialization.}
\label{fig:scenario3}
\end{figure}

\subsection{Overhead of Generated Delta Code}\label{sec:overhead}
\inverda's delta code is generated from Datalog rules and aims at a general and solid solution.
So far, our focus is on the correct propagation of data access on multiple co-existing schema versions.
We expect the database optimizer to find a fast execution plan, however, there will be an overhead of \inverda compared to hand-optimized SQL.

\textbf{\taskyX:} In Figure~\ref{fig:taskyOverhead}, we use the previously presented \taskyX example with \num{100000} tasks and compare the performance of \inverda generated delta code to the handwritten one.
There are two aspects to observe.
First, the hand-optimized delta code causes slightly less (up to \SI{4}{\%}) overhead than the generated one.
Considering the difference in length and complexity of the code (\SI{359}{x} LOC for the evolution), a performance overhead of \SI{4}{\%} in average is more than reasonable for most users.
Second, the materialization significantly influences the actual performance.
Reading the data in the materialized version is up to twice as fast as accessing it from the respective other version in this scenario.
For the write workload (insert new tasks), we observe again a reasonably small overhead compared to handwritten SQL.
Interestingly, the evolved materialization is always faster because the initial materialization requires to manage an additional auxiliary table for the foreign key relationship.
A DBA can optimize the overall performance for a given workload by adapting the materialization, which is a very simple task with \inverda.
An advisor tool supporting the optimization task is very well imaginable, but out of scope for this paper.

\begin{figure*}
\begin{subfigure}{.33\textwidth}
\includegraphics[width=\columnwidth, page=1, trim=0mm 11mm 0mm 20mm, clip=true]{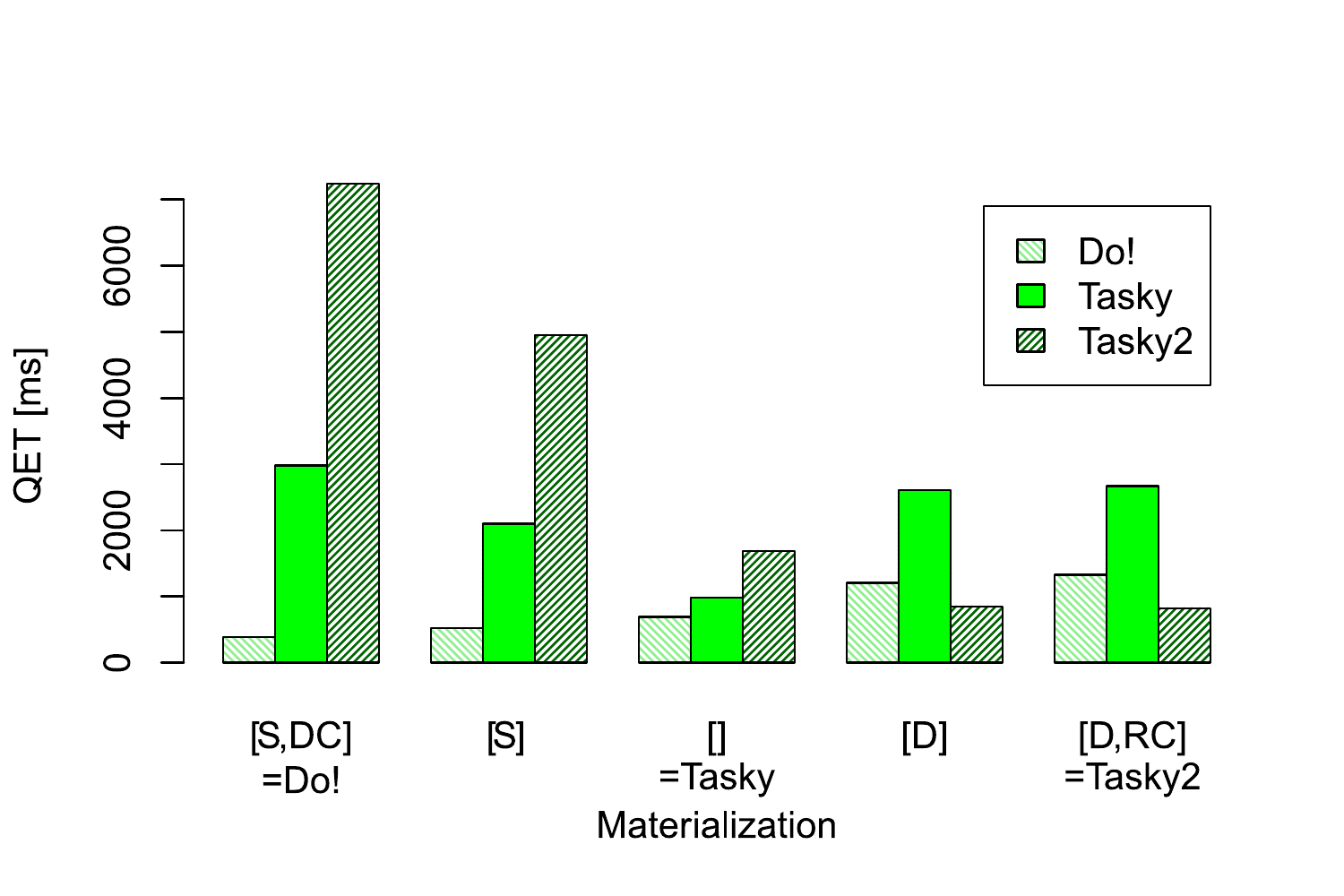}\vspace{-1mm}
\caption{Materializations for \taskyX Mix.}
\label{fig:taskyMix}
\end{subfigure}
\begin{subfigure}{.33\textwidth}
\includegraphics[width=\columnwidth, page=1, trim=0mm 11mm 0mm 20mm, clip=true]{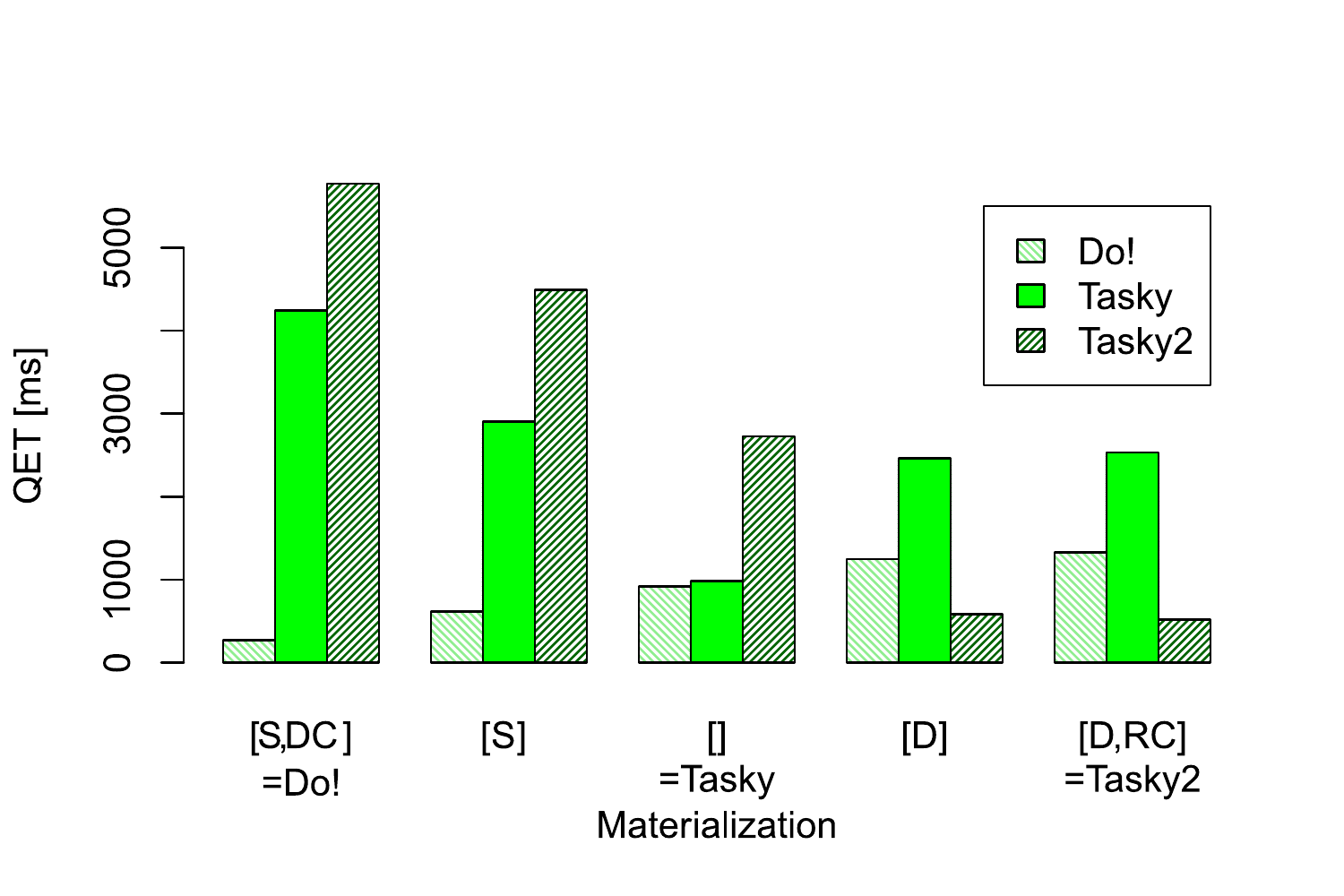}\vspace{-1mm}
\caption{Materializations for \taskyX Read.}
\label{fig:taskyRead}
\end{subfigure}
\begin{subfigure}{.32\textwidth}
\includegraphics[width=\columnwidth, page=1, trim=0mm 11mm 0mm 20mm, clip=true]{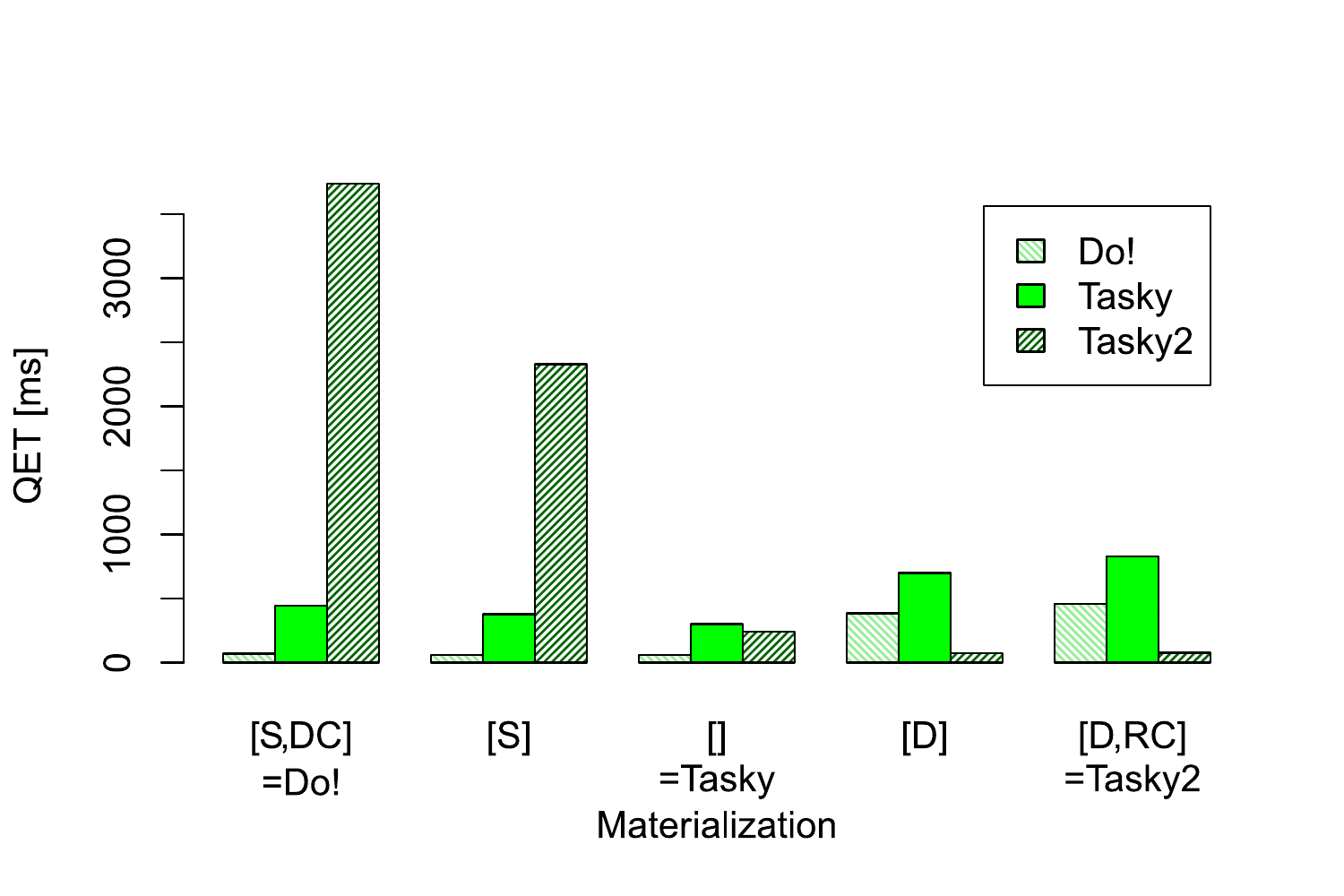}\vspace{-1mm}
\caption{Materializations for \taskyX Write.}
\label{fig:taskyWrite}
\end{subfigure}
\vspace{-3mm}
\caption{Different workloads on all possible materialization of \taskyX.}
\label{fig:tasky}
\end{figure*}

\subsection{Benefit of Flexible Materialization}\label{sec:independence}
Adapting the physical table schema to the current workload is hard with handwritten SQL, but almost for free with \inverda (\num{1} LOC instead of \num{182} in our \taskyX example).
Let's assume a development team spares the effort for rewriting delta code and works with a fixed materialization.

\textbf{\taskyX:} Again, we use the \taskyX example with \num{100000} tasks.
Figure~\ref{fig:scenario2} shows the accumulated propagation overhead for handwritten SQL with the two fixed materializations and for \inverda with an adaptive materialization.
Assume, over time the workload changes from \SI{0}{\%} access to \taskyyX and \SI{100}{\%} to \taskyX to the opposite \SI{100}{\%} and \SI{0}{\%} according to the Technology Adoption Life Cycle.
The adoption is divided into \num{1000} time slices where \num{1000} queries are executed respectively.
The workload mixes \SI{50}{\%} reads, \SI{20}{\%} inserts, \SI{20}{\%} updates, and \SI{10}{\%} deletes.
As soon as the evolved materialization is faster for the current workload mix, we instruct \inverda to change the materialization.
As can be seen, \inverda facilitates significantly better performance---including migration cost---than a fixed materialization.

This effect increases with the length of the evolution, since \inverda can also materialize intermediate stages of the evolution history.
Assume, all users use exclusively the mobile phone app \dodoX; but as \taskyyX gets released users switch to \taskyyX which comes with its own mobile app.
In Figure~\ref{fig:scenario3}, we simulate the accumulated overhead for either materializing one of the three schema versions or for a flexible materialization.
The latter starts at \dodoX, moves to \taskyX after several users started using \taskyyX, and finally moves to \taskyyX when the majority of users did so.
Again, \inverda's flexible materialization significantly reduces the overhead for data propagation without any interaction of a developer.

The DBA can choose between multiple materialization schemas.
The number of valid materialization schemas greatly depends on the actual structure of the evolution.
The lower bound is a linear sequence of depending SMOs, e.g. one table with $N$ \sql{ADD COLUMN} SMOs has $N$ valid materializations.
The upper bound are $N$ independent SMOs, each evolving another table, with $2^N$ valid materializations.
Specifically, the \taskyX example has five valid materializations.

Figure~\ref{fig:tasky} shows the data access performance on the three schema versions for each of the five materialization schema.
The materialization schemas are represented as the lists of SMOs that are materialized.
We use abbreviations for SMOs: e.g. $[D,RC]$ on the very right corresponds to schema version \taskyyX since both the decompose SMO (D) and the rename column SMO (RC) are materialized.
The initial materialization is in the middle, while e.g. the materialization according to \dodoX is on the very left.
The workload mixes \SI{50}{\%} reads, \SI{20}{\%} inserts, \SI{20}{\%} updates, \SI{10}{\%} deletes in Figure~\ref{fig:taskyMix}, \SI{100}{\%} reads in Figure~\ref{fig:taskyRead}, and \SI{100}{\%} inserts in Figure~\ref{fig:taskyWrite} on the depicted schema versions.
Again, the measurements show that accesses to each schema version are fastest when its respective table versions are materialized, i.e. when the physical table schema fits the accessed schema version.
However, there are differences in the actual overhead, so the globally optimal materialization depends on the workload distribution among the schema version.
E.g. writing to \taskyyX is \num{49} times faster when the physical table schema matches \taskyyX instead of \dodoX.
This gain increases with every SMO, so for longer evolutions with more SMOs it will be even higher.

\begin{figure}[b]
\centering
\includegraphics[width=\columnwidth, page=1, trim=0mm 6mm 0mm 6mm, clip=true]{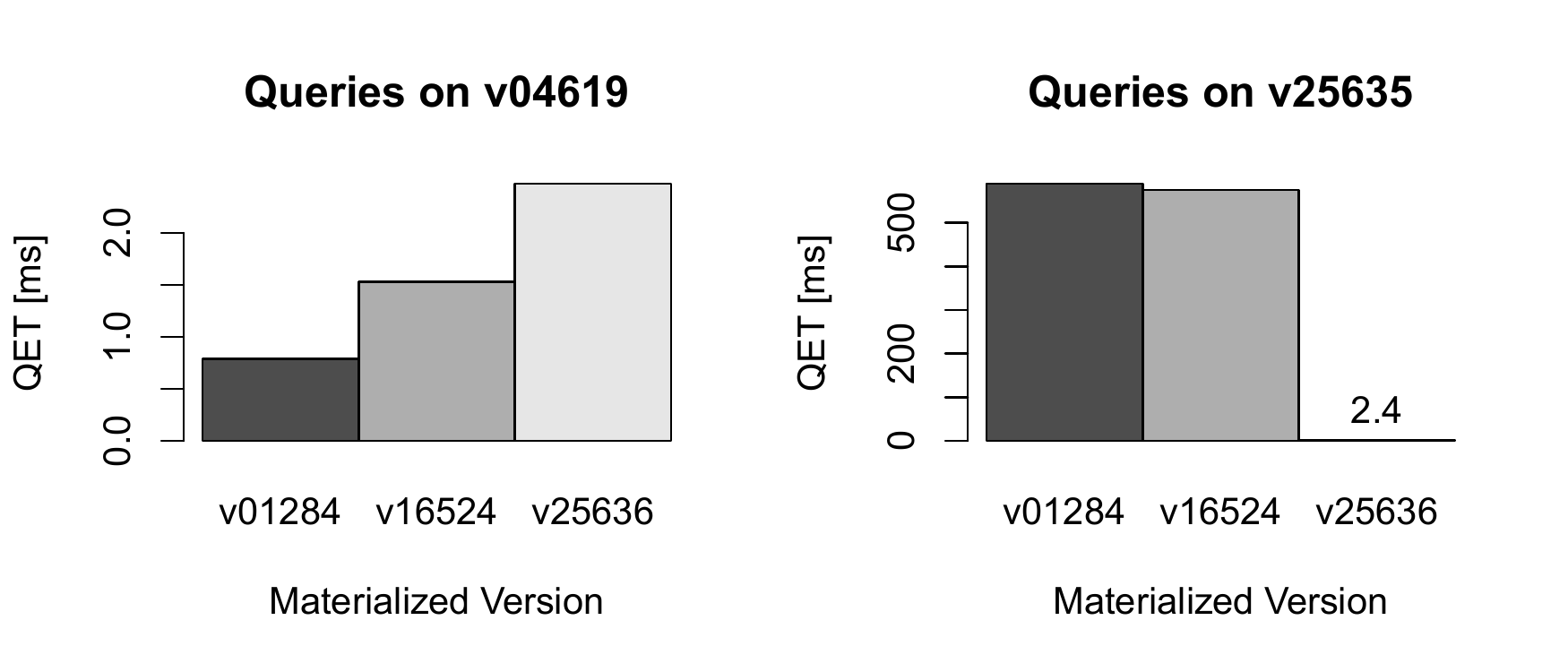}
\caption{Optimization potential for Wikimedia.}
\label{fig:wikiPerformance}
\end{figure}

\textbf{Wikmedia:} 
The benefits of the flexible materialization originate from the increased performance when accessing data locally without the propagation through SMOs.
We load our Wikimedia with the data of Akan Wiki in schema version \emph{v16524} (109th version) with \num{14359} pages and \num{536283} links.
We measure the read performance for the template queries from~\cite{CarloA.Curino} both in schema version \emph{v04619} (28th version) and \emph{v25635} (171th version).
The chosen materializations match version \emph{v01284} (1st), \emph{v16524} (109th), and \emph{v25635} (171th) respectively.
In Figure~\ref{fig:wikiPerformance}, a great performance difference of up to two orders of magnitude is visible, so there is a huge optimization potential.
We attribute this asymmetry to the dominance of add column SMOs, which need an expensive join with an auxiliary table to propagate data forwards, but only in a cheap projection to propagate backwards.

\begin{figure*}
\includegraphics[width=\textwidth, page=1, trim=0mm 22mm 5mm 25mm, clip=true]{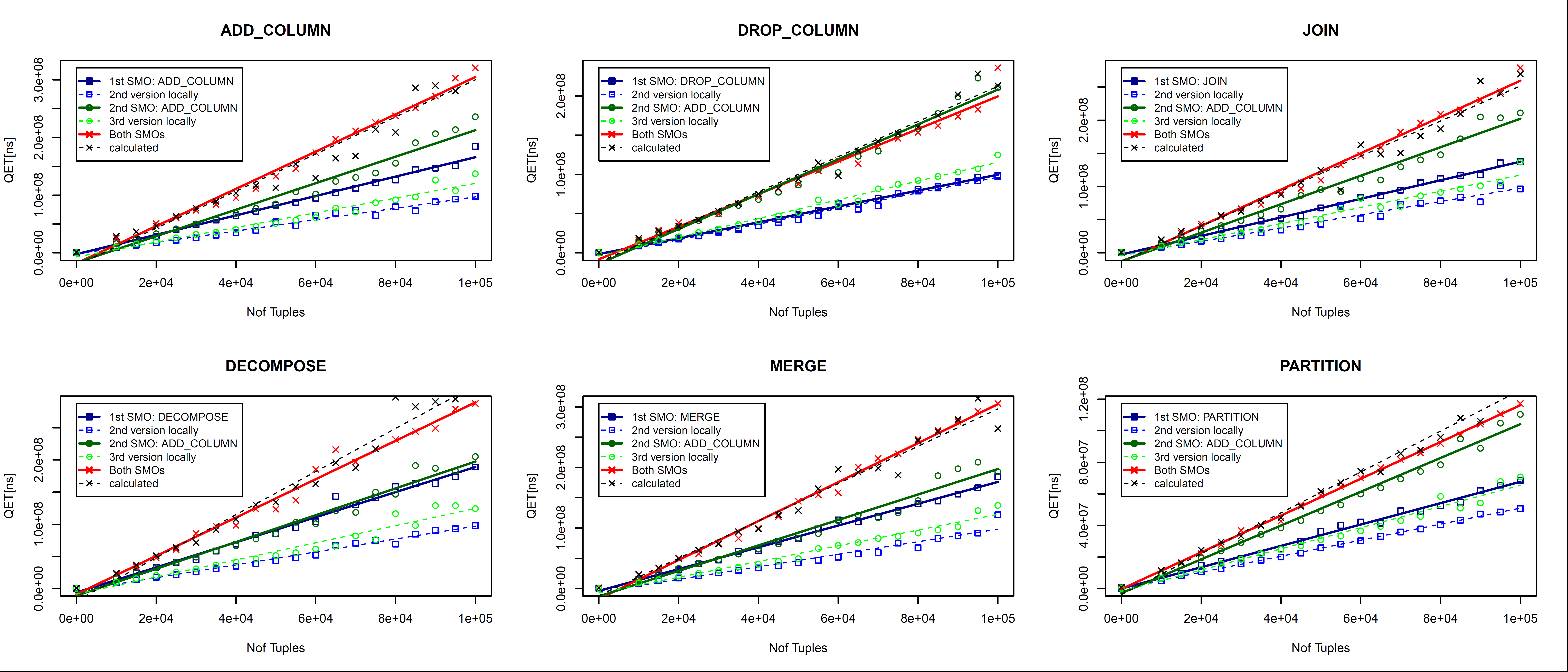}
\caption{Scaling behavior of the ADD COLUMN SMO.}
\label{fig:add_column}
\end{figure*}

\textbf{All possible evolutions with two SMOs:}
To show that it is always possible to gain a better performance by optimizing the materialization, we conduct a micro benchmark on all possible evolutions with two SMOs---except of creating and dropping tables as well as renaming columns and tables, since they have no relevant performance overhead in the first place.
We show that there is always a performance benefit when accessing data locally compared to propagating it through SMOs and we disprove that \inverda might add complexity to the data access, so two SMOs do not impact each other negatively.
We generate evolutions with two SMOs and three schema versions: 1st version -- 1st SMO -- 2nd version -- 2nd SMO -- 3rd version.
The second version always contains a table $R(a,b,c)$; the number of generated tuples in this table is the x-axis of the charts.
In Figure~\ref{fig:add_column}, we exemplarily consider all the combinations with add column as 2nd SMO, since this is the most common one.
Again, accessing data locally is up to twice as fast as propagating it through an SMO, so the optimization potential exists in all scenarios.
The average speedup over all SMOs is \SI{2.1}{}.
We calculate the expected performance for the combination of both SMOs as the sum of both query execution times minus reading data locally at the 2nd schema version.
This is reasonable since the data for the 2nd SMO is already in memory after executing the 1st one.
Figure~\ref{fig:add_column} shows that the measured time for propagating the data through two SMOs is always in the same range as the calculated combination of the overhead of the two SMOs individually, so we showed that there is great optimization potential for all combinations of those SMOs and we can safely use it without fearing additional overhead when combining SMOs.
This holds for all pairs of SMOs: on average the measured time differs only \SI{6.3}{\%} from the calculated one. 
In sum, \inverda enables the DBA to easily adapt the materialization schema to a changing workload and to significantly speed up query processing without hitting other stakeholders' interests. 

\section{Related Work}\label{sec:rw}
Both the database evolution~\cite{Rahm2006} and co-existing schema versions~\cite{Roddick} are well recognized in database research.
For database evolution, existing approaches increase comfort and efficiency, for instance by defining a schema evolution aware query language~\cite{Roddick1992} or by providing a general framework to describe database evolution in the context of evolving applications~\cite{Dominguez2008}.
With Meta Model Management 2.0~\cite{Bernstein2007b}, Phil Bernstein et al. introduced a comprehensive tooling to i.a. match, merge, and diff given schema versions.
The resulting mappings couple the evolution of both the schema and the data just as our SMOs do; however, the difference is that mappings are derived from given schema versions while \inverda takes developer-specified mappings and derives the new schema version.
Currently, PRISM~\cite{Curino2012} appears to provide the most advanced database evolution tool with an SMO-based DEL.
PRISM was introduced in 2008 and focused on the plain database evolution~\cite{Curino2008}.
Later, PRISM++ added constraint evolution and update rewriting~\cite{Curino2012}.

These existing works provide a great basis for database evolution and \inverda builds upon them to add \textbf{logical data independence} and \textbf{co-existence of schema versions}, which basically requires bidirectional transformations~\cite{Terwilliger2012a}.
Particularly, symmetric relational lenses lay a foundation to describe read and write accesses along a bidirectional mapping~\cite{Hofmann2011}.
For \inverda, we adapt this idea to bidirectional SMOs.
Another extension of PRISM++ takes a first step towards co-existing schema versions by answering queries on former schema versions w.r.t.\ to the current data~\cite{Moon2009}, however, \inverda also covers write operations on those former schema versions.
There are multiple systems also taking this step, however, the DELs are usually rather limited or work on different meta models like data warehouses~\cite{key:article}.
The ScaDaVer system~\cite{Wall2011} allows additive and subtractive SMOs on the relational model, which simplifies bidirectionality and hence it is a great starting point towards more powerful DELs.
\indel also covers restructuring SMOs and is based on established DELs~\cite{Curino2012, HVBL15}.
To the best of our knowledge, we are the first to realize end-to-end support for co-existing schema versions based on such powerful DELs.

\section{Conclusions}\label{sec:conclusion}
Current DBMSes do not support co-existing schema versions properly, forcing developers to manually write complex and error-prone delta code, which propagates read/write accesses between schema versions.
Moreover, this delta code needs to be adapted manually whenever the DBA changes the physical table schema.
\inverda greatly simplifies creating and maintaining co-existing schema versions for developers, while the DBA can freely change the physical table schema.
For this sake, we have introduced \indel, an intuitive bidirectional database evolution language that carries enough information to generate all the delta code automatically.
We have formally validated \indel's bidirectionality making it a sound and robust basis for \inverda.
In our evaluation, we have shown that \indel scripts are significantly shorter than handwritten SQL scripts (\SI{359}{x}).
The performance overhead caused by the automatically generated delta code is very low but the freedom to easily change the physical table schema is highly valuable: we can greatly speed up query processing by matching the physical table schema to the current workload.
In sum, \inverda finally enables agile---but also robust and maintainable---software development for information systems.
Future research topics are (1) zero-downtime migrations, (2) efficient physical table schemas e.g. with redundancy, (3) self-managed DBMSes continuously adapting the physical table schema to the current workload, and (4) optimized delta code within a database system instead of triggers.


\bibliographystyle{abbrv}
\newpage
\begin{minipage}{\columnwidth}
\bibliography{bib}
\end{minipage}


\newpage
\appendix
\section{Bidirectionality of Split}
In this paper (Section~\ref{sec:bidirectionality}), we merely showed one of the two bidirectionality conditions for the \sql{SPLIT} SMO to explain the concept.
As a reminder, the two conditions are:
\begin{align}
\dt&=\getTarget^{data}(\getSource(\dt))\label{ap:dt}\\
\ds&=\getSource^{data}(\getTarget(\ds))\label{ap:ds}
\end{align}
We have already shown Condition~\ref{ap:ds}, so we do the same for Condition~\ref{ap:dt}, now.

Writing data \Rdata and \Sdata from the target-side to the source-side is done with the mapping $\getSource(\Rdata, \Sdata)$.
With source-side materialization all target-side auxiliary tables are not required, so we apply Lemma~\ref{lemma:empty} to obtain:
\begin{align}
\mathbf{\getSource(\Rdata, \Sdata):}&\nonumber\\
T(\pk,A)\leftarrow& \Rdata(\pk,A)\\
T(\pk,A)\leftarrow& \Sdata(\pk,A), \neg \Rdata(\pk,\_)\\
R^-(\pk) \leftarrow& \Sdata(\pk,A),\neg \Rdata(\pk,\_),\cR (A)\\
R^*(\pk) \leftarrow& \Rdata(\pk,A),\neg \cR(A)\\
S^+(\pk,A)\leftarrow& \Sdata(\pk,A), \Rdata(\pk,A'),A\neq A'\\
S^-(\pk) \leftarrow& \Rdata(\pk,A),\neg \Sdata(\pk,\_),\cS (A)\\
S^*(\pk) \leftarrow& \Sdata(\pk,A),\neg \cS(A)
\end{align}
Reading the target-side data back from the source-side adds the rule set \getTarget(Rule~\ref{rule:getrg:begin}--\ref{rule:getrg:end}) to the mapping.
Using Lemma~\ref{lemma:deducation}, the mapping $\getTarget(\getSource(\Tdata))$ extends to:

\begin{align}
\mathbf{\getTarget(\getSource}&\mathbf{(\Rdata, \Sdata)):}\nonumber\\
R(\pk,A)\leftarrow& \Rdata(\pk,A),\cR (A), \neg \Sdata(\pk,\_)\\
R(\pk,A)\leftarrow& \Rdata(\pk,A),\cR (A), \Rdata(\pk,\_)\\
R(\pk,A)\leftarrow& \Rdata(\pk,A),\cR (A), \Sdata(\pk,A'), \neg \cR (A')\\
R(\pk,A)\leftarrow& \Rdata(\pk,A), \Rdata(\pk,A),\neg \cR(A)\\
\nonumber\\[-3mm]
R(\pk,A)\leftarrow& \mathbf{\Sdata(\pk,A)}, \neg \Rdata(\pk,\_),\cR (A), \mathbf{\neg \Sdata(\pk,A)}\\
R(\pk,A)\leftarrow& \Sdata(\pk,A), \mathbf{\neg \Rdata(\pk,\_)},\cR (A), \mathbf{\Rdata(\pk,\_)}\\
R(\pk,A)\leftarrow& \Sdata(\pk,A), \neg \Rdata(\pk,\_),\mathbf{\cR (A)}, \mathbf{\neg \cR (A)}\\
R(\pk,A)\leftarrow& \Sdata(\pk,A), \mathbf{\neg \Rdata(\pk,\_)}, \mathbf{\Rdata(\pk,A)},\neg \cR(A)\\
\nonumber\\
S(\pk,A)\leftarrow& \mathbf{\Rdata(\pk,A)},\cS (A), \mathbf{\neg \Rdata(\pk,A)}, \neg \Sdata(\pk,\_)\\
S(\pk,A)\leftarrow& \mathbf{\Rdata(\pk,A)},\cS (A), \mathbf{\neg \Rdata(\pk,A)}, \neg \Rdata(\pk,\_)\\
S(\pk,A)\leftarrow& \mathbf{\Rdata(\pk,A)},\cS (A), \mathbf{\neg \Rdata(\pk,A)}, \nonumber\\& \Sdata(\pk,A'), \Rdata(\pk,A),A'=A\\
S(\pk,A)\leftarrow& \Rdata(\pk,A),\cS (A), \mathbf{\Sdata(\pk,\_)}, \mathbf{\neg \Sdata(\pk,\_)}\\
S(\pk,A)\leftarrow& \mathbf{\Rdata(\pk,A)},\cS (A), \Sdata(\pk,\_), \mathbf{\neg \Rdata(\pk,\_)}\\
S(\pk,A)\leftarrow& \Rdata(\pk,A),\cS (A), \Sdata(\pk,\_), \nonumber\\& \Sdata(\pk,A'), \Rdata(\pk,A),A'=A\label{ap:so}\\
S(\pk,A)\leftarrow& \Rdata(\pk,A),\mathbf{\cS (A)}, \mathbf{\neg \cS(A)}, \neg \Sdata(\pk,\_)\\
S(\pk,A)\leftarrow& \Rdata(\pk,A),\mathbf{\cS (A)}, \mathbf{\neg \cS(A)}, \neg \Rdata(\pk,A)\\
S(\pk,A)\leftarrow& \Rdata(\pk,A),\mathbf{\cS (A)}, \mathbf{\neg \cS(A)}, \nonumber\\& \Sdata(\pk,A'), \Rdata(\pk,A),A'=A
\end{align}

\begin{align}
S(\pk,A)\leftarrow& \mathbf{\Sdata(\pk,A)}, \neg \Rdata(\pk,\_),\cS (A), \neg \Rdata(\pk,\_), \nonumber\\& \mathbf{\neg \Sdata(\pk,\_)}\\
S(\pk,A)\leftarrow& \Sdata(\pk,A), \neg \Rdata(\pk,\_),\cS (A), \neg \Rdata(\pk,\_), \nonumber\\& \neg \Rdata(\pk,\_)\\
S(\pk,A)\leftarrow& \Sdata(\pk,A), \mathbf{\neg \Rdata(\pk,\_)},\cS (A), \neg \Rdata(\pk,\_), \nonumber\\& \Sdata(\pk,A), \mathbf{\Rdata(\pk,A'')},A=A''\\
S(\pk,A)\leftarrow& \mathbf{\Sdata(\pk,A)}, \neg \Rdata(\pk,\_),\cS (A), \Sdata(\pk,\_), \nonumber\\& \mathbf{\neg \Sdata(\pk,\_)}\\
S(\pk,A)\leftarrow& \Sdata(\pk,A), \neg \Rdata(\pk,\_),\cS (A), \Sdata(\pk,\_), \nonumber\\& \neg \Rdata(\pk,\_)\\
S(\pk,A)\leftarrow& \Sdata(\pk,A), \mathbf{\neg \Rdata(\pk,\_)},\cS (A), \Sdata(\pk,\_), \nonumber\\& \Sdata(\pk,A), \mathbf{\Rdata(\pk,A'')},A=A''\\
S(\pk,A)\leftarrow& \mathbf{\Sdata(\pk,A)}, \neg \Rdata(\pk,\_),\cS (A), \Rdata(\pk,A'),\nonumber\\& \cS (A'), \mathbf{\neg \Sdata(\pk,\_)}\\
S(\pk,A)\leftarrow& \Sdata(\pk,A), \mathbf{\neg \Rdata(\pk,\_)},\cS (A), \mathbf{\Rdata(\pk,A')},\nonumber\\& \cS (A'), \neg \Rdata(\pk,\_)\\
S(\pk,A)\leftarrow& \Sdata(\pk,A), \mathbf{\neg \Rdata(\pk,\_)},\cS (A), \mathbf{\Rdata(\pk,A')},\nonumber\\& \cS (A'), \Sdata(\pk,A), \Rdata(\pk,A''),A=A''\\
\nonumber\\[-3mm]
S(\pk,A)\leftarrow& \Sdata(\pk,A), \Rdata(\pk,A'),A\neq A'\\
\nonumber\\[-3mm]
S(\pk,A)\leftarrow& \Rdata(\pk,A), \mathbf{\Sdata(\pk,A)},\neg \cS(A), \mathbf{\neg \Sdata(\pk,\_)}\\
S(\pk,A)\leftarrow& \mathbf{\Rdata(\pk,A)}, \Sdata(\pk,A),\neg \cS(A), \mathbf{\neg \Rdata(\pk,\_)}\\
S(\pk,A)\leftarrow& \Rdata(\pk,A), \Sdata(\pk,A),\neg \cS(A), \nonumber\\& \Sdata(\pk,A), \Rdata(\pk,A),A=A\\
S(\pk,A)\leftarrow& \mathbf{\Sdata(\pk,A)}, \neg \Rdata(\pk,\_), \Sdata(\pk,A),\neg \cS(A), \nonumber\\& \mathbf{\neg \Sdata(\pk,\_)}\\
S(\pk,A)\leftarrow& \Sdata(\pk,A), \neg \Rdata(\pk,\_), \Sdata(\pk,A),\neg \cS(A), \nonumber\\& \neg \Rdata(\pk,\_)\\
S(\pk,A)\leftarrow& \Sdata(\pk,A), \mathbf{\neg \Rdata(\pk,\_)}, \Sdata(\pk,A),\neg \cS(A), \nonumber\\& \Sdata(\pk,A), \mathbf{\Rdata(\pk,A')},A=A')
\\
T'(\pk,A)\leftarrow& \mathbf{\Rdata(\pk,A)},\neg \cR (A),\neg \cS (A), \mathbf{\neg  \Rdata(\pk,\_)}, \nonumber\\& \neg  \Sdata(\pk,\_)\\
T'(\pk,A)\leftarrow& \mathbf{\Rdata(\pk,A)},\neg \cR (A),\neg \cS (A), \mathbf{\neg  \Rdata(\pk,\_)}, \nonumber\\& \Sdata(\pk,A'), \cS(A')\\
T'(\pk,A)\leftarrow& \Rdata(\pk,A),\mathbf{\neg \cR (A)},\neg \cS (A), \Rdata(\pk,A), \nonumber\\& \mathbf{\cR(A)}, \neg  \Sdata(\pk,\_)\\
T'(\pk,A)\leftarrow& \Rdata(\pk,A),\mathbf{\neg \cR (A)},\neg \cS (A), \Rdata(\pk,A), \nonumber\\& \mathbf{\cR(A)}, \Sdata(\pk,A''), \cS(A'')\\
\nonumber\\[-3mm]
T'(\pk,A)\leftarrow& \mathbf{\Sdata(\pk,A)}, \neg \Rdata(\pk,\_),\neg \cR (A),\neg \cS (A), \nonumber\\& \neg  \Rdata(\pk,\_), \mathbf{\neg  \Sdata(\pk,\_)}\\
T'(\pk,A)\leftarrow& \Sdata(\pk,A), \neg \Rdata(\pk,\_),\neg \cR (A),\mathbf{\neg \cS (A)}, \nonumber\\& \neg  \Rdata(\pk,\_), \Sdata(\pk,A), \mathbf{\cS(A)})\\
T'(\pk,A)\leftarrow& \mathbf{\Sdata(\pk,A)}, \neg \Rdata(\pk,\_),\neg \cR (A),\neg \cS (A), \nonumber\\& \Rdata(\pk,A'), \cR(A'), \mathbf{\neg  \Sdata(\pk,\_)}\\
T'(\pk,A)\leftarrow& \Sdata(\pk,A), \mathbf{\neg \Rdata(\pk,\_)},\neg \cR (A),\neg \cS (A), \nonumber\\& \mathbf{\Rdata(\pk,A')}, \cR(A'), \Sdata(\pk,A''), \cS(A'')
\end{align}

Using Lemma~\ref{lemma:unsatisfiability} we remove all rules that have contradicting literals (marked bold).
Particularly, there remains no rule for $T'$ as expected.
Further, we remove duplicate literals within the rules, so we obtain the simplified rule set:
\begin{align}
R(\pk,A)\leftarrow& \Rdata(\pk,A),\cR (A), \neg \Sdata(\pk,\_)\label{ap:r1}\\
R(\pk,A)\leftarrow& \Rdata(\pk,A),\cR (A)\label{ap:r2}\\
R(\pk,A)\leftarrow& \Rdata(\pk,A),\cR (A), \Sdata(\pk,A'), \neg \cR (A')\label{ap:r3}\\
R(\pk,A)\leftarrow& \Rdata(\pk,A),\neg \cR(A)\label{ap:r4}\\
\nonumber\\
S(\pk,A)\leftarrow& \Sdata(\pk,A), \Rdata(\pk,A),\cS (A)\label{ap:s0}\\
S(\pk,A)\leftarrow& \Sdata(\pk,A), \neg \Rdata(\pk,\_),\cS (A)\label{ap:s1}\\
S(\pk,A)\leftarrow& \Sdata(\pk,A), \neg \Rdata(\pk,\_),\cS (A)\label{ap:s2}\\
S(\pk,A)\leftarrow& \Sdata(\pk,A), \Rdata(\pk,A'),A\neq A'\label{ap:s3}\\
S(\pk,A)\leftarrow& \Sdata(\pk,A), \Rdata(\pk,A), \neg \cS(A)\label{ap:s4}\\
S(\pk,A)\leftarrow& \Sdata(\pk,A), \neg \Rdata(\pk,\_), \neg \cS(A)\label{ap:s5}
\end{align}
Rule~\ref{ap:s0} is derived from Rule~\ref{ap:so} by applying the equivalence of $A$ and $A'$ to the remaining literals.
Let's now focus on the rules for $R$.
Rules~\ref{ap:r1} and~\ref{ap:r3} are subsumed by Rule~\ref{ap:r2}, since they contain the identical literals as Rule~\ref{ap:r2} plus additional conditions.
Lemma~\ref{lemma:tautology} allows us to further reduce Rules~\ref{ap:r2} and~\ref{ap:r4}, so we achieve that all tuples in $R$ survive one round trip without any information loss or gain:
\begin{align}
R(\pk,A)\leftarrow& \Rdata(\pk,A)
\end{align}
We also reduce the rules for $S$.
Rule~\ref{ap:s2} can be removed, since it is equal to Rule~\ref{ap:s1}.
With Lemma~\ref{lemma:tautology}, Rules~\ref{ap:s1} and~\ref{ap:s5} as well as Rules~\ref{ap:s0} and~\ref{ap:s4} can be combined respectively.
This results in the following rules for $S$:
\begin{align}
S(\pk,A)\leftarrow& \Sdata(\pk,A), \Rdata(\pk,A)\label{ap:s10}\\
S(\pk,A)\leftarrow& \Sdata(\pk,A), \neg \Rdata(\pk,\_)\label{ap:s11}\\
S(\pk,A)\leftarrow& \Sdata(\pk,A), \Rdata(\pk,A'),A\neq A'\label{ap:s12}
\end{align}
Rules~\ref{ap:s10} and~\ref{ap:s12} basically state that the payload data in $R$ ($A$ and $A'$ respectively) is either equal to or different from the payload data in $S$ for the same key $\pk$.
When we rewrite Rule~\ref{ap:s10} to:
\begin{align}
S(\pk,A)\leftarrow& \Sdata(\pk,A), \Rdata(\pk,A'), A=A'
\end{align}
we can apply Lemma~\ref{lemma:tautology} to obtain the two rules:
\begin{align}
S(\pk,A)\leftarrow& \Sdata(\pk,A), \Rdata(\pk,\_)\\
S(\pk,A)\leftarrow& \Sdata(\pk,A), \neg \Rdata(\pk,\_)
\end{align}
With the help of Lemma~\ref{lemma:tautology}, we reduce $\getTarget(\getSource(\Rdata, \Sdata))$ to
\begin{align}
R(\pk,A)\leftarrow& \Rdata(\pk,A)\\
S(\pk,A)\leftarrow& \Sdata(\pk,A) \qed
\end{align}
So, both Condition~\ref{ap:dt} and Condition~\ref{ap:ds} for the bidirectionallity of the split SMO are formally validated by now.
Since the merge SMO is the inverse of the split SMO and uses the exact same mapping rules vice versa, we also implicitly validated the bidirectionality of the merge SMO.
We can now safely say: wherever we materialize the data, the mapping rules always guarantee that each table version can be accessed just like a regular table---no data will be lost or gained.
This is a strong guarantee and the basis for \inverda to provide co-existing schema versions within a single database.

\section{Remaining SMO{\normalsize \textbf{s}}}\label{apx:SMOs}
We introduce the syntax and semantics of the remaining SMOs from Figure~\ref{fig:smos}.
Further, we include the results of the formal evaluation of their bidirectionality.
Please note that creating, dropping, and renaming tables as well as renaming columns exclusively affects the schema version catalog and does not include any kind of data evolution, hence there is no need to define mapping rules for these SMOs.
In Section~\ref{ap:ac}, we introduce the \sql{ADD COLUMN} SMO.
Since SMOs are bidirectional, exchanging the rule sets \getSource and \getTarget yields the inverse SMO: \sql{DROP COLUMN}.
There are different extends for the \sql{JOIN} and its inverse \sql{DECOMPOSE} SMO: a join can have inner or outer semantics and it can be done based on the primary key, a foreign key, or on an arbitrary condition.
As summarized in Table~\ref{tab:decomposeAndJoin}, each configuration requires different mapping functions, however some are merely the inverse or variants of others.
The inverse of \sql{DECOMPOSE} is \sql{OUTER JOIN} and joining at a foreign key is merely a specific condition.

\subsection{Add Column / Drop Column}\label{ap:ac}
\noindent\textbf{SMO:} \sql{\sqlk{ADD COLUMN} $b$ \sqlk{AS} \tbl{$f$}{$r_1$,\ldots,$r_n$} \sqlk{INTO} $R$} \\
\textbf{Inverse:} \sql{\sqlk{DROP COLUMN} $b$ \sqlk{FROM} $R$ \sqlk{DEFAULT} \tbl{$f$}{$r_1$,\ldots,$r_n$}}\\
The \sql{ADD COLUMN} SMO adds a new column $b$ to a table $R$ and calculates the new values for $b$ according to the given function $f$.
The inverse \sql{DROP COLUMN} SMO uses the same parameters to ensure bidirectionality.
\begin{align}
\mathbf{\getTarget:}\hspace{3mm} R'(\pk,A,b)&\leftarrow R(\pk,A),b=f_B(\pk,A),\neg B(\pk,\_)\\
R'(\pk,A,b)&\leftarrow R(\pk,A), B(\pk,b)\\
\mathbf{\getSource:}\hspace{6.7mm}R(\pk,A)&\leftarrow R'(\pk,A,\_)\\
B(\pk,b)&\leftarrow R'(\pk,\_,b)\\[3mm]
\noalign{$\mathbf{\getSource(\getTarget(\Rdata)):}$}
R(\pk,A)&\leftarrow \Rdata(\pk,A)\\
B(\pk,b)&\leftarrow \Rdata(\pk,A),b=f_B(\pk,A) \qed\label{align:ac}\\
\noalign{$\mathbf{\getTarget(\getSource(\Rdata')):}$}
R'(\pk,A,b)&\leftarrow \Rdata'(\pk,A,b) \qed
\end{align}
The auxiliary table $B$ stores the values of the new column when the SMO is virtualized to ensure bidirectionality.
With the projection to data tables, the SMO satisfies Conditions~\ref{ap:dt} and~\ref{ap:ds}.
For repeatable reads, the table $B$ is also needed when data is given in the source schema version (Rule~\ref{align:ac}).

\subsection{Decompose on the Primary Key}\label{sec:decomposePK}
\noindent\textbf{SMO:} \sql{\sqlk{DECOMPOSE TABLE} $R$ \sqlk{INTO} \tbl{$S$}{$A$}, \tbl{$T$}{$B$} \sqlk{ON} \sqlk{PK}}\\
\textbf{Inverse:} \sql{\sqlk{OUTER} \sqlk{JOIN TABLE} $S$, $T$ \sqlk{INTO} $R$ \sqlk{ON} \sqlk{PK}}\\
To fill the gaps potentially resulting from the inverse outer join, we use the null value $\omega_R$.
The bidirectionality conditions are satisfied: after one round trip, no data is lost or gained.
\begin{align}
\mathbf{\getTarget:}\hspace{22mm} S(\pk,A) &\leftarrow R(\pk,A,\_), A\neq \omega_R\\
T(\pk,B) &\leftarrow R(\pk,\_,B), B\neq\omega_R\\
\mathbf{\getSource:}\hspace{17.5mm} R(\pk,A,B)&\leftarrow S(\pk,A), T(\pk,B)\\
R(\pk,A,\omega_R)&\leftarrow S(\pk,A), \neg T(\pk,\_)\\
R(\pk,\omega_R,B)&\leftarrow \neg S(\pk,\_), T(\pk,B)\\[3mm]
\mathbf{\getSource(\getTarget(\Rdata)):}\hspace{2mm} R(\pk,A,B)&\leftarrow \Rdata(\pk,A,B) \qed\\
\mathbf{\getTarget(\getSource(\Sdata, \Tdata)):} S(\pk,A) &\leftarrow \Sdata(\pk,A)\\
T(\pk,B) &\leftarrow \Tdata(\pk,B) \qed
\end{align}

\begin{table}
\centering\scriptsize
\begin{tabular}{lccc}
\toprule
&\textbf{Decompose}&\textbf{Outer Join}&\textbf{Inner Join}\\
\midrule
\textbf{ON PK}&\emph{\ref{sec:decomposePK}}&Inverse of \ref{sec:decomposePK}&\emph{\ref{sec:joinPK}}\\
\textbf{ON FK}&\emph{\ref{sec:decomposeFK}}&Inverse of \ref{sec:decomposeFK}&Variant of \ref{sec:joinCond}\\
\textbf{ON Cond.}&\emph{\ref{sec:decomposeCond}}&Inverse of \ref{sec:decomposeCond}&\emph{\ref{sec:joinCond}}\\
\bottomrule
\end{tabular}
\caption{Overview of different Decompose and Join SMOs.}
\label{tab:decomposeAndJoin}
\end{table}

\subsection{Decompose on a Foreign Key}\label{sec:decomposeFK}
\noindent\textbf{SMO:} \sql{\sqlk{DECOMPOSE TABLE} $R$ \sqlk{INTO} \tbl{$S$}{$A$}, \tbl{$T$}{$B$} \sqlk{ON} \sqlk{FK} $t$}\\
\textbf{Inverse:} \sql{\sqlk{OUTER} \sqlk{JOIN TABLE} $S$, $T$ \sqlk{INTO} $R$ \sqlk{ON} \sqlk{FK} $t$}\\
A \sql{DECOMPOSE}, which creates a new foreign key, needs to generate new identifiers.
Assume we cut away the addresses from persons stored in one table, we eliminate all duplicates in the new address table, assign a new identifier to each address, and finally add a foreign key column to the new persons table.
On every call, the function $id_T(B)$ returns a new unique identifier for the payload data $B$ in table $T$.
In our implementation, this is merely a regular SQL sequence and the mapping rules ensure that an already generated identifier is reused for the same data.
In order to guarantee proper evaluation of these functions, we enforce a sequential evaluation of the rules by distinguishing between existing and new data.
For a literal $L$, we use the indexes $L_{o}$ (old) and $L_{n}$ (new) to note the difference, however they have now special semantics and are evaluated like any other literal in Datalog.
For a \sql{DECOMPOSE ON FK}, we propose the rule set:
\begin{align}
\noalign{$\mathbf{\getTarget:}$}
T_{n}(t,B)&\leftarrow R(\pk,\_,B),ID_R(\pk,t)\\
T_{n}(t,B)&\leftarrow R(\pk,\_,B),\neg ID_R(\pk,t),\nonumber\\&\phantom{\leftarrow} \neg T_{o}(\_,B),t=id_T(B)\label{align:dc4}\\
T_{n}(t,B)&\leftarrow R(\_,\_,B),T_{o}(t,B)\label{align:dc3}\\
S(\pk,A,t) &\leftarrow R(\pk,A,\_), ID_R(\pk,t)\\
S(\pk,A,\omega) &\leftarrow R(\pk,A,\_), ID_R(\pk,\omega)\\
S(\pk,A,t) &\leftarrow R(\pk,A,B), \neg ID_R(\pk,\_), T_{n}(t,B)\label{align:dc5}\\
\noalign{$\mathbf{\getSource:}$}
R(\pk,A,B)&\leftarrow S(\pk,A,t), T(t,B)\\
R(\pk,A,\omega)&\leftarrow S(\pk,A,\omega)\\
R(t,\omega,B)&\leftarrow \neg S(\_,\_,t), T(t,B)\\
ID_R(\pk,t)&\leftarrow S(\pk,\_,t), T(t,\_)\\
ID_R(\pk,\omega)&\leftarrow S(\pk,\_,\omega)\\
ID_R(t,t)&\leftarrow \neg S(\_,\_,t), T(t,\_)\\[3mm]
\noalign{$\mathbf{\getSource(\getTarget(\Rdata)):}$}
R(\pk,A,B)&\leftarrow \Rdata(\pk,A,B)\\
ID_R(\pk,t)&\leftarrow \Rdata(\pk,A,B),t=id_T(B) \qed\\
\noalign{$\mathbf{\getTarget(\getSource(\Sdata, \Tdata)):}$}
T(t,B)&\leftarrow \Tdata(t,B)\\
S(\pk,A,t) &\leftarrow \Sdata(\pk,A,t) \qed
\end{align}
Projecting the outcomes to the data tables, again satisfies our bidirectionality Conditions~\ref{ap:dt} and~\ref{ap:ds}.
Hence, no matter whether the SMO is virtualized or materialized, both the source and the target side behave like common single-schema databases.
Storing data in $R$ implicitly generates new values to the auxiliary table $ID_R$, which is intuitive: we need to store the assigned identifiers for the target version to ensure repeatable reads on those generated identifiers.

\subsection{Decompose on Condition}\label{sec:decomposeCond}
\noindent\textbf{SMO:} \sql{\sqlk{DECOMPOSE TABLE} $R$ \sqlk{INTO} \tbl{$S$}{$A$}, \tbl{$T$}{$B$} \sqlk{ON} $\cond(A,B)$}\\
\textbf{Inverse:} \sql{\sqlk{OUTER} \sqlk{JOIN TABLE} $S$, $T$ \sqlk{INTO} $R$ \sqlk{ON} $\cond(A,B)$}\\
To e.g. normalize a table that holds books and authors ($N:M$), we can either use two subsequent \sql{DECOMPOSE ON FK} to maintain the relationship between books and authors, or---if the new evolved version just needs the list of authors and the list of books---we simply split them giving up the relationship.
In the following, we provide rules for the latter case.
Either way we have to generate new identifiers for both the books and the authors.
We use the same identity generating function as in Section~\ref{sec:decomposeFK}.
\begin{align}
\noalign{$\mathbf{\getTarget:}$}
S_{n}(s,A) \leftarrow& R(r,A,\_), ID_{o}(r,s,\_)\\
S_{n}(s,A) \leftarrow& R(r,A,\_), \neg ID_{o}(r,\_,\_),\nonumber\\& A\neq\omega_R, s=id_S(A)\\
S_{n}(r,A) \leftarrow& R(r,A,\_), \neg ID_{o}(r,\_,\_), A=\omega_R\\
T_{n}(t,B)\leftarrow& R(r,\_,B),ID_{o}(r,\_,t)\\
T_{n}(t,B)\leftarrow& R(r,\_,B),\neg ID_{o}(r,\_,\_),\nonumber\\& B\neq\omega_R,t=id_T(B)\\
T_{n}(r,B) \leftarrow& R(r,\_,B),\neg ID_{o}(r,\_,\_), B=\omega_R\\
ID_{n}(r,s,t) \leftarrow& R(r,A,B),S_{n}(s,A),T_{n}(t,B)\\
R^-(s,t)\leftarrow& \neg R(\_,A,B),S_{n}(s,A),\nonumber\\&T_{n}(t,B), \cond(A,B)\\
\noalign{$\mathbf{\getSource:}$}
R_{o}(r,A,B)\leftarrow& S(s,A), T(t,B), ID_{o}(r,s,t)\\
R_{o}(r,A,B)\leftarrow& S(s,A), T(t,B), \cond(A,B), \neg R^-(s,t), \nonumber\\& \neg ID_{o}(\_,s,t),r=id_R(A,B)\\
ID_{n}(r,s,t)\leftarrow& S(s,A), T(t,B), \cond(A,B),R_{o}(r,A,B)\\
ID_{n}(r,s,t)\leftarrow& ID_{o}(r,s,t)\\
R_{n}(r,A,B)\leftarrow& R_{o}(r,A,B)\\
R_{n}(s,A,\omega_R)\leftarrow&S(s,A),\neg ID_{n}(\_,s,\_)\\
R_{n}(t,\omega_R,B)\leftarrow&T(t,B),\neg ID_{n}(\_,\_,t)\\[3mm]
\noalign{$\mathbf{\getSource(\getTarget(\Sdata, \Tdata)):}$}
R_{n}(r,A,B) \leftarrow& \Rdata(r,A,B)\\
ID_{n}(r,s,t)\leftarrow& \Rdata(r,A,B), \nonumber\\&s=id_S(A),t=id_T(B) \qed\\
\noalign{$\mathbf{\getTarget(\getSource(\Rdata)):}$}
S(s,A)\leftarrow& \Sdata(s,A)\\
T(t,B)\leftarrow& \Tdata(t,B)\\
ID(r,s,t) \leftarrow& \Sdata(s,A),\Tdata(t,B), \nonumber\\&\cond(A,B),r=id_R(A,B)\qed
\end{align}
The bidirectionality Conditions~\ref{ap:dt} and~\ref{ap:ds} are satisfied.
For repeatable reads, the auxiliary table $ID$ stores the generated identifiers independently of the chosen materialization.

\subsection{Inner Join on Primary Key}\label{sec:joinPK}
\noindent\textbf{SMO:} \sql{\sqlk{JOIN TABLE} $R$, $S$ \sqlk{INTO} $T$ \sqlk{ON} \sqlk{PK}}\\
For this join, we merely need one auxiliary table to store those tuples that do not match with a join partner.
Since both bidirectionality conditions hold in the end, we have formally shown the bidirectionality of \sql{JOIN ON PK}.

\begin{align}
\mathbf{\getTarget:}\hspace{18mm} R(\pk,A,B)&\leftarrow S(\pk,A), T(\pk,B)\\
S^+(\pk,A)&\leftarrow S(\pk,A), \neg T(\pk,\_)\\
T^+(\pk,B)&\leftarrow \neg S(\pk,\_), T(\pk,B)\\
\mathbf{\getSource:}\hspace{22mm} S(\pk,A) &\leftarrow R(\pk,A,\_)\\
S(\pk,A) &\leftarrow S^+(\pk,A)\\
T(\pk,B) &\leftarrow R(\pk,\_,B)\\
T(\pk,B) &\leftarrow T^+(\pk,B)\\[3mm]
\mathbf{\getSource(\getTarget(\Sdata, \Tdata)):}S(\pk,A) &\leftarrow \Sdata(\pk,A)\\
T(\pk,B) &\leftarrow \Tdata(\pk,B) \qed \\
\mathbf{\getTarget(\getSource(\Rdata)):}\hspace{1.5mm} R(\pk,A,B)&\leftarrow \Rdata(\pk,A,B) \qed
\end{align}

\subsection{Inner Join on Condition}\label{sec:joinCond}
\noindent\textbf{SMO:} \sql{\sqlk{JOIN TABLE} $R$, $S$ \sqlk{INTO} $T$ \sqlk{ON} $\cond(A,B)$}\\
A join on a condition creates new tuples, so we have to generate new identifiers as well.
We use the notion introduced in Section~\ref{sec:decomposeFK} and satisfy the bidirectionality conditions.
\begin{align}
\noalign{$\mathbf{\getTarget:}$}
R_{n}(r,A,B)\leftarrow& S(s,A), T(t,B), ID_{o}(r,s,t)\\
R_{n}(r,A,B)\leftarrow& S(s,A), T(t,B), \cond(A,B),\neg R^-(s,t), \nonumber\\& \neg ID_{o}(\_,s,t),r=id_R(A,B)\\
ID_{n}(r,s,t)\leftarrow& S(s,A), T(t,B), \cond(A,B),R_{n}(r,A,B)\\
ID_{n}(r,s,t)\leftarrow& ID_{o}(r,s,t)\\
S^+(s,A)\leftarrow& S(s,A), \neg ID_{n}(\_,s,\_)\\
T^+(t,B)\leftarrow& T(t,B), \neg ID_{n}(\_,\_,t)\\
\noalign{$\mathbf{\getSource:}$}
S_{n}(s,A) \leftarrow& R(r,A,\_), ID(r,s,\_)\\
S_{n}(s,A) \leftarrow& R(r,A,\_), \neg ID(r,s,\_), s=id_S(A)\\
S_{n}(s,A) \leftarrow& S^+(s,A)\\
T_{n}(t,B)\leftarrow& R(r,\_,B),ID(r,\_,t)\\
T_{n}(t,B)\leftarrow& R(r,\_,B),\neg ID(r,\_,t),t=id_T(B)\\
T_{n}(t,B) \leftarrow& T^+(t,B)\\
ID(r,s,t) \leftarrow& R(r,A,B),S_{n}(s,A),T_{n}(t,B)\\
R^-(s,t)\leftarrow& \neg R(\_,A,B),S_{n}(s,A),T_{n}(t,B), \cond(A,B)\\[3mm]
\noalign{$\mathbf{\getSource(\getTarget(\Sdata, \Tdata)):}$}
S(s,A)\leftarrow& \Sdata(s,A)\\
T(t,B)\leftarrow& \Tdata(t,B)\\
ID(r,s,t) \leftarrow& \Sdata(s,A),\Tdata(t,B), \nonumber\\&\cond(A,B),r=id_R(A,B)\qed\\
\noalign{$\mathbf{\getTarget(\getSource(\Rdata)):}$}
R_{n}(r,A,B) \leftarrow& \Rdata(r,A,B)\\
ID_{n}(r,s,t)\leftarrow& \Rdata(r,A,B), \nonumber\\&s=id_S(A),t=id_T(B) \qed
\end{align}

In sum, \indel's SMOs are formally guaranteed to be bidirectional: a solid ground for co-existing schema versions.

\balance
\end{document}